\documentclass[1p]{elsarticle}

\usepackage{fancyhdr}
\usepackage{booktabs}
\usepackage{ctable}
\usepackage{amsmath}
\usepackage{mathtools}
\usepackage{commath}
\usepackage{graphicx}

\usepackage{graphics}
\usepackage{tabularx}
\usepackage{color}
\usepackage{geometry}
\usepackage{natbib}
\usepackage{mciteplus}
\usepackage{footnote}
\usepackage{amssymb}
\usepackage{pdflscape}
\usepackage{afterpage}
\usepackage{multirow}
\usepackage[title]{appendix}
\usepackage{subfigure}
\usepackage{caption}
\usepackage{lineno}
\usepackage{amssymb}
\usepackage{natbib}
\usepackage{mleftright} 
\usepackage{latexsym}
\usepackage{color,soul}
\usepackage[intoc]{nomencl}
\usepackage{array}
\newcolumntype{P}[1]{>{\centering\arraybackslash}p{#1}}
\makenomenclature
 
\usepackage{etoolbox}
\renewcommand\nomgroup[1]{%
	\item[\bfseries
	\ifstrequal{#1}{A}{}{%
		\ifstrequal{#1}{G}{Greek Letters}{%
			\ifstrequal{#1}{S}{Subscripts}{}}}%
	]}

\usepackage[version=3]{mhchem} 

\biboptions{numbers,sort&compress}

\begin{document}
\begin{frontmatter}

\title{Computational fluid dynamics (CFD) simulation of three\textendash phase non\textendash Newtonian slurry flows in industrial horizontal pipelines}


\author[1]{Mohsen Sadeghi}
\author[1]{Somasekhara Goud Sontti}

\address[1]{Department of Chemical and Materials Engineering, University of Alberta, Alberta T6G 1H9, Canada}
\author[2]{Enzu Zheng}
\address[2]{CSIRO Mineral Resources , Clayton, VIC 3168, Australia}


\author[1]{Xuehua Zhang\corref{cor1}}
\ead{Xuehua.Zhang@ualberta.ca}
\cortext[cor1]{Corresponding author}

\begin{abstract}
\noindent 

Understanding the flow behavior of complex concentrated slurries is of tremendous importance for industrial waste management. In this study, the transport of three\textendash phase oil sands tailings in a horizontal pipeline is simulated via the mixture multiphase model coupled with the kinetic theory of granular flow. The solid particles and bitumen droplets are conveyed via a non\textendash Newtonian carrier fluid in a turbulent regime inside an industrial\textendash scale pipeline. The simulation results showed exceptional agreement with the field data, with errors of $<$\,3.5\% for velocity distribution and $<$\,15\% for the pressure drop. A systematic parametric investigation was performed for a wide range of flow conditions, showing that the majority of bitumen droplets reside at the top region of the pipe. Our findings may help design an effective process for the separation of bitumen residues during pipeline transport. 

\end{abstract}


\end{frontmatter}

\newpage
\section{Introduction}

Concentrated slurry flows are frequently encountered in many industries such as minerals extraction \citep{pullumanrev}, food and farms \citep{food}, and petroleum \citep{ZAMBRANO2017}. Transport of extraction or production wastes, i.e., tailings in horizontal pipelines is one example of where the slurry flow management is relevant. A slurry flow is a multiphase system where solid particles are transported via water as the carrier. In reality, the solid particles in the slurry are different in size, also called as polydisperse particles \citep{pullumanrev, LI2018915}. The presence of fine particles, i.e., particles with sizes less than 44\,$\mu$m can alter the rheological behavior of the carrier by forming a pseudo\textendash homogeneous mixture with water. The carrier fluid in slurry transport usually behaves as a non\textendash Newtonian fluid, possessing a yield stress and a decrease in viscosity as the shear rate increases, i.e., shear\textendash thinning fluid \citep{enzu2021,pullumanrev}. Considering the relevant flow velocities in the industry, and the pipeline characteristics, the transport process may occur in a turbulent regime. A turbulent slurry flow can be classified as a highly complex system, and development of a predictive model is of great importance for pipeline design and optimization of the transport process.  

Several groups of researchers have conducted experimental, modeling, and simulation works to understand the slurry flow characteristics and optimize the process with various configurations and applications \citep{Mishra_2019}. However, these type of studies have been performed mostly for two\textendash phase slurries. Experimental investigations in pilot and lab\textendash scale pipelines have been thoroughly reported in the literature for both Newtonian \citep{gillies1991,gillies1994,gillies2000,gillies2004,KAUSHAL2005,KAUSHAL2007} and non\textendash Newtonian \citep{enzu2021,EESA2008997,eesa2009,ignatenko2018continual} carrier fluids. For a Newtonian carrier fluid, the researchers at the Saskatchewan Research Council have reported experimental data for slurry flows in various systems and different flow conditions \citep{gillies1994,gillies1991,gillies2000,gillies2004}, they extensively investigated the concentration distribution of coarse particles and determined the effect of flow parameters such as mixture velocity on the solids concentration distribution. Moreover, experimental data of concentration distribution and pressure drop has been presented by \citet{KAUSHAL2005} and and \citet{KAUSHAL2007} for flow velocities up to 5\,m/s and solids volume fraction up to 0.5, where they identified the maximum concentration for smaller particles closer to the bottom region, but closer to the core for coarse particles. Furthermore, \citet{Vlasak2020} reported the pressure gradient and solids concentration distribution for turbulent flow inside inclined pipes, with flow velocities up to 2.07\,m/s and solids volume fraction up to 0.34, where they identified insignificant changes in solids concentration distribution with relatively small changes in pipe inclination, e.g., less than 30°.  

For the non\textendash Newtonian carrier fluid, recently \citet{penik} and \citet{enzu2021} reported the concentration distribution and pressure drop in a two\textendash phase flow with a carrier fluid fitted with a Herschel\textendash Bulkley model in a turbulent regime. The solids volume fractions were up to 0.2 and 0.16, and flow velocities were up to 4.51 and 3.67\,m/s in the works by \citet{penik}, and \citet{enzu2021}, respectively, and more symmetrical solids concentration distributions in case of higher mixture velocities were identified in both works. \citet{matouvsek2015experimental} used a Herschel\textendash Bulkely carrier fluid to drive a solid bed to sliding and measured the pressure gradient and mean delivered solids concentration for flow velocities up to 4.76\,m/s, and confirmed the suitability of a proposed two\textendash layer model for the prediction of viscous frictions in a laminar regime. Moreover, \citet{ignatenko2018continual} reported only the pressure drop data for the Herschel\textendash Bulkley carrier fluid in a turbulent regime with flow velocities up to 5.2\,m/s and particles concentration of 10\%. 

\citet{turian1998flow} presented experimental data of pressure gradient for both laminar and turbulent slurry flows with solids volume fraction up to 0.252 where the carrier fluid follows the Sisko rheological model. Studying the effect of mixture velocity on the pressure gradient is the common feature of the works by \citet{ignatenko2018continual}, and \citet{turian1998flow}, while the evolution of pressure gradient in full transition of laminar to turbulent flow was reported. For a power\textendash law carrier fluid, \citet{eesa2009} measured the solid\textendash phase velocity distribution in a laminar flow with outlet solids volume fraction up to 0.4 and flow velocities up to 125\,mm/s. In their earlier study, \citet{EESA2008997} reported the velocity distribution in a similar system and flow conditions but with an Ellis carrier fluid, where they reported more symmetrical solid velocity distributions in the case of smaller solid particles and higher mixture velocities and solids concentration.  


 As complement to experimental studies, the development of reliable models validated through a broad range of flow conditions is of great importance for understanding and predication of the slurry flow behavior. The attempts to simulate the slurry flow behavior started with the theoretical models and empirical and semiempirical correlations. \citet{gillies1994} conducted systematic experiments to measure the concentration distribution of coarse particles in radial direction, and proposed a method to predict the concentration distribution neglecting the particles interactions. In their other studies, they proposed an approach to predict the head loss called the two\textendash layer model originated by \citet{wilson76} and extended it to various conditions \citep{gillies1991,gillies2000}. In \citet{wilson76}'s model, a stratified solid\textendash liquid flow has been modeled assuming two separate layers in which the mixture was considered in single phase having uniform velocity and solid concentration. 
Similarly, other groups of researchers have developed Two\textendash Layer \citep{gavignet,Doron1987,rojas} and Three\textendash Layer models \citep{doron1993three,RAMADAN2005}.


In parallel with the efforts to improve experimental results and analytical models, computational fluid dynamics (CFD) is becoming more comprehensive to investigate the basic fundamental understanding of  slurry flows in a pipeline. Different multiphase flow modeling approaches such as Eulerian\textendash Eulerian \citep{embarka,sadeghi,akaushal2012,zhang2021}, Eulerian\textendash Lagrangian \citep{ZHOU2020,chen2020,desjardins}, and mixture model \citep{Liu2019,ignatenko2018continual,ZAMBRANO2017} have been utilized. Recently, \citet{MESSA2020} and \citet{ALOBAID2022} presented a comprehensive description and comparison of the multiphase modeling approaches.

In the Eulerian\textendash Eulerian (or Two\textendash Fluid) model, the phases are treated as interpenetrating continua, and governing equations are applied to each phase while the phases can exchange momentum via interphase forces. The kinetic theory of granular flow  is usually coupled with the Eulerian\textendash Eulerian model to capture the effects of particles interactions on the flow field. This approach makes the consideration of chemical reactions and mass transfer between the phases possible, and offers a relatively good compromise between the computational cost and accuracy \citep{ALOBAID2022}. In the Eulerian\textendash Lagrangian scheme, the particles, droplets or bubbles (based on the type of the secondary phase) are tracked individually, and Newton's law of motion is applied to each particle. This approach is especially applicable when the concentration of the secondary phase is dilute, otherwise the computational cost will be tremendous due to the high number of equations to be solved \citep{messarev,ALOBAID2022}. The mixture model is a simplified Eulerian multiphase model that can be beneficial in modeling of complex multiphase systems , e.g., when the number of secondary phases is high or the interactions between the secondary phases are not well\textendash known\citep{fluent2011ansys}. 

 The treatment of the multiphase mixture and lower number of differential equations to be solved, make the mixture model an efficient method for the systems with polydisperse particles and multiple phases \citep{messarev}. Multiphase slurry systems, cyclone separators \citep{PADHI2019115698,KOU2020cyclone}, cavitation processes \citep{SHI2019672}, and nanofluidics \citep{REZAEI2018nano} are some examples of where the mixture model can be applied and has been used previously by other researchers. For the multiphase slurry systems, although numerous efforts have been made to develop models capable of predicting the flow behavior, the lack of a model which can capture the non-Newtonian behavior of the carrier and effects of particle size distribution (PSD) is sensed. Notably, the effects of an additional secondary phase and mutual effects of other phases and flow conditions on the additional phase is another aspect that is not studied in the open literature.\\

 The oil sands tailings slurry is a byproduct of bitumen extraction, a mixture with three phases consisting polydisperse solid particles ( about 55 wt\% ), water, and a minute fraction of residual bitumen (about 1 wt\%) \citep{zhourev,tailings}. The oil sands tailings are transported to large open areas called tailings ponds and stored so the particles settle, the water is recycled, and the land could be reclaimed. However, the residual bitumen in the slurry is not treated and separated, and can pose substantial threats to the environment, including decomposition into greenhouse gases and contribution to the global warming \citep{amir}. In this regard, separation of bitumen from the tailings before disposal to the tailings ponds is crucial. Understanding the behavior and the effective parameters of the tailings flow is a significant step toward designing a separation technology. Nevertheless, the complexity of the mixture and the low fraction of bitumen in the slurry makes the separation process difficult. There is no predictive model for the transport characteristics of slurry flow in large-diameter pipes, especially for multiple secondary phase solids along with bitumen droplets. Importantly, the effects of PSD and non-Newtonian viscosity have not been studied thoroughly in a complex multiphase flow system. Therefore, in depth understanding of slurry flow characteristics and operating conditions are important to design a suitable process. \\

In the present study, a turbulent three-phase tailings slurry flow in an industrial-scale pipeline was successfully modeled using the mixture multiphase model coupled with the KTGF. The flow consists of highly concentrated polydisperse solid particles, a non-Newtonian carrier fluid and a small fraction of bitumen droplets. The model was validated against ten sets of field data of pressure gradient and velocity distribution. Systematic parametric study was performed to investigate the effect of mixture velocity, particle size distribution, non-Newtonian viscosity, and pipe angle on the flow behavior. To the best of our knowledge, the present study is a  first-of-a-kind on highly concentrated slurry transport in a industrial pipeline. The results of this study would contribute to better understanding of three-phase slurry system and would be used for pipeline design and optimization, as well as a guideline for designing a separation process inside the pipeline.

\section{Methodology}

\subsection{Collection of field data}
\label{expsection}

\begin{figure}[!ht]
\centering

  \includegraphics[width=0.99\textwidth]{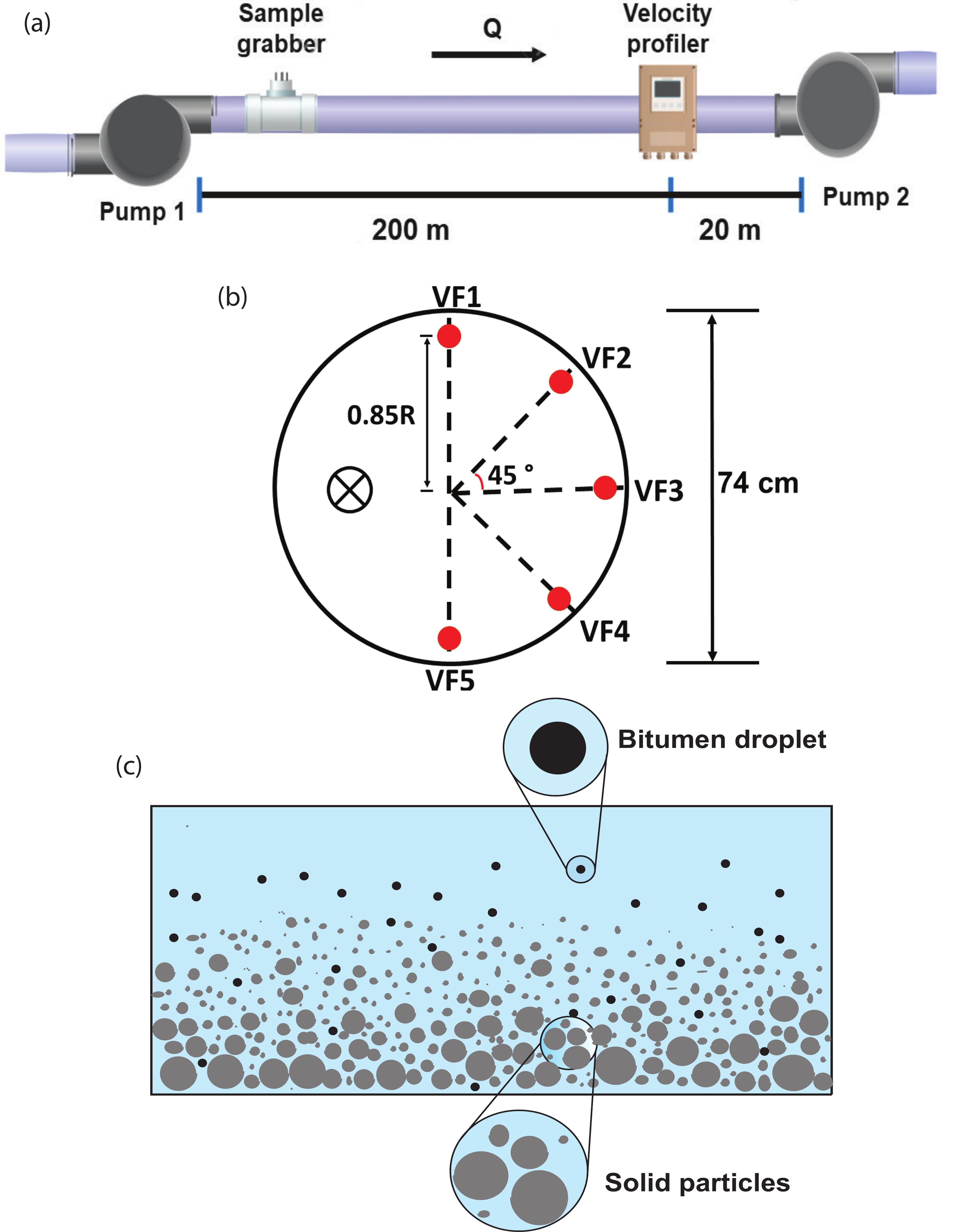}
\caption{Schematics of (a) the tailings hydrotransport pipeline, (b) the velocity profiler with the tangential and radial locations of the measurement points (The flow direction is into the plane and has been marked), and (c) sketch of a tailings slurry flow in a horizontal pipeline .}\label{fig:pipeline}
\end{figure}

Fig.\,\ref{fig:pipeline}\,a shows the schematic of the pipeline where the measurements have been performed to collect field data that are used for model validation. This pipeline is a section of a mining operation. The pipe is 74\,cm in diameter, and located horizontally with 2 pumps and measurement devices in\textendash line. The section used to measure the flow conditions for this study is around 220\,m in length. The solid composition  samples are  collected after the first pump discharge for every 12 hours and  and the compositions of the mixture are determined using a Dean\textendash Stark apparatus \citep{dean}. The mixture  velocities are measured approximately 20\,m distance from the  the second pump suction. A sieving method is used to calculate the particles size distribution for the solid particles. The pressures were measured at the discharge of the first pump and suction of the second pump using two gauge pressure meter devices. To calculate the pressure gradient, the average values of pressure at the first pump discharge and second pump suction are considered.

For the velocity measurements, a velocity profiler is used. Fig.\,\ref{fig:pipeline}b shows the schematic view of the measurement locations, from VF1 at the top to VF5 at the bottom of the pipe. The velocities are measured using a SANDtrac Velocity Profile System (CiDRA) based on a non\textendash invasive technique. This device is an array of five sensors attached to the pipe wall, which track the turbulent eddies inducing pressure disturbance and consequently forces on the wall. This array senses the dynamic strains applied to the pipe by these forces and converts them to electrical signals to be interpreted, and the velocity values are calculated \citep{sonartrac}. This technique is well\textendash established and applied commercially for slurry transport in the industry, where detection of the formation of stationary beds is a usage of the device \citep{cidra1,cidra2}. The distance where the velocities are measured from the pipe center is around 0.85R which R is the pipe radius. Measurements of velocity and flow rate are taken every two seconds, yielding multiple values for a given period with some fluctuations.

Fig. \ref{fig:pipeline}\,c shows a schematic representation of the tailings flow with polydisperse solid particles and bitumen droplets in the pipeline. The data of compositions (mass fractions of different components of the flow) and PSD are collected every 12 hours. For the simulation purposes, multiple time windows of 30 minutes which contain 900 time steps in the measurements are selected, and the flow conditions have been averaged over the whole or a portion of these time windows. In the data sets collected, some fluctuations in terms of flow conditions are observed, which is reasonable considering the highly turbulent flow inside an industrial\textendash scale pipeline. To remove the effect of these fluctuations, averaging is needed. This procedure is explained in detail in section\,\ref{section:expresults}.


\subsection{Mixture model}

\begin{table}[hbt!]
	\centering
	\abovedisplayskip=0pt
	\belowdisplayskip=0pt
	\caption{Mass and momentum equations for the mixture model employed in the present CFD simulation.}\label{tab:CFD_model_equations}
	\begin{tabular}{ m{2cm} m{10.5cm} }\hline

		Continuity \newline\newline & {\begin{subequations}\begin{flalign}
					&\frac{\partial}{\partial t}\left(\rho_{m}\right)+\nabla \cdot\left(\rho_{m} \vec{V}_{m}\right)=0 &\\
					&\vec{V}_{m}=\frac{\sum_{k=1}^{n} \alpha_{k} \rho_{k} \vec{V}_{k}}{\rho_{m}}&
				\end{flalign}\label{eq:continuity}\end{subequations}}\\[-2ex]

		Momentum  \newline\newline \newline\newline \newline\newline \newline & {\begin{subequations}\begin{flalign}
					& \begin{multlined}
						\frac{\partial}{\partial t}\left(\rho_{m} \vec{V}_{m}\right)+\nabla \cdot\left(\rho_{m} \vec{V}_{m} \vec{V}_{m}\right)=-\nabla p+\nabla \cdot\left[\overline{\bar{\tau}}_{s}\right] +\rho_{m} \vec{g}+\vec{F}\\+\nabla \cdot\left(\sum_{k=1}^{n} \alpha_{k} \rho_{k} \vec{V}_{dr, k} \vec{V}_{dr, k}\right)
					\end{multlined}&\\
					&\overline{\bar{\tau}}_{s}= \mu_{m}\left(\nabla \vec{V}_{m}+\nabla \vec{V}_{m}^{T}\right) &\\
					&\vec{V}_{dr,p}=\vec{V}_{pq}-\sum_{k=1}^{n} c_{k} \vec{V}_{q k} &\\
					&\vec{V}_{pq}=\frac{\left(\rho_{p}-\rho_{m}\right) d_{p}^{2}}{18 \mu_{q} f_{drag}} \vec{a}-\frac{\eta_{t}}{\sigma_{t}}\left(\frac{\nabla a_{p}}{a_{p}}-\frac{\nabla a_{q}}{a_{q}}\right) &\\
					&\frac{\partial}{\partial t}\left(\alpha_{p} \rho_{p}\right)+\nabla \cdot\left(\alpha_{p} \rho_{p} \vec{V}_{m}\right)= 0&
				\end{flalign}\label{eq:momentum}\end{subequations}}\\[-2ex]

		\hline\end{tabular}
\end{table}

In this section, the theory and equations of the mixture multiphase model are provided. The flow field of the mixture in the pipeline is predicated by solving the conservation of mass and momentum equations in three\textendash dimensional Cartesian coordinates using mixture model \citep{manninen} based on finite volume method. The continuity and momentum balance equations are written and solved only for the mixture phase, with an algebraic expression for the slip velocity between the primary and secondary phases. 

The equations for conservation of mass and momentum are given in Table \ref{tab:CFD_model_equations}. The continuity equation for the mixture is given as Eq.\,\eqref{eq:continuity}a
where $\vec{V}_{m}$ is the mass\textendash averaged velocity expressed as Eq.\,\eqref{eq:continuity}b,
and $\rho_{m}$ is the density of the mixture. In Eq.\,\eqref{eq:continuity}b, $\alpha_{k}$ and $\rho_{k}$ denote the volume fraction and density of $k^{th}$ phase, respectively.


In the mixture model, the momentum balance equation is written by summing the individual momentum equations for all phases as Eq.\,\eqref{eq:momentum}a. In this equation, $n$ is the number of phases, $\vec{F}$ is a body force, and $\mu_{m}$ is the viscosity of the mixture ($\mu_{m}=\sum_{k=1}^{n} a_{k} \mu_{k}$).
$\vec{V}_{dr, k}$ is the drift velocity for secondary phase $k$, and is defined as the relative velocity between $k$ phase
velocity and the velocity of center of mass. 
The slip velocity ($V_{pq}$) is defined as the velocity of a secondary phase $(p)$ relative to the velocity of the primary phase $(q)$, and is connected to the drift velocity as Eq.\,\eqref{eq:momentum}c, where $c_{k}=\frac{\alpha_{k} \rho_{k}}{\rho_{m}}$. \citet{manninen} mixture model assumes a local equilibrium between the phases over a short spatial length scale, and uses an algebraic slip formulation.

 In a turbulent regime, the slip velocity can be expressed as Eq.\,\eqref{eq:momentum}d. In this equation, $\tau_{p}$ is the particle relaxation time
$\tau_{p}=\frac{\rho_{p} d_{p}^{2}}{18{\mu_{q}}}$, $d$ is the diameter of the particles (or droplets or bubbles) of secondary phase $p, \vec{a}$ is the secondary phase particle's acceleration. $\sigma_{t}$ is a Prandtl/Schmidt number set to 0.75 and $\eta_{t}$ is the turbulent diffusivity. 

The drag functions $f_{d r a g}$ are taken  from  \citet{gidaspow1991hyd} (fluid\textendash solid) and symmetric (fluid\textendash fluid) model between carrier and secondary phases. For the liquid\textendash solid drag model (the drag model for the carrier fluid and solids), numerous studies in the literature show that the model proposed by \citet{gidaspow1991hyd} gives an accurate expression \citep{sadeghi}. Nevertheless, as the number of studies for the three\textendash phase simulation with two fluid phases is limited, no reliable analysis for the choice of drag model between the carrier and bitumen can be found in the literature. In this regard, the available drag models in Ansys Fluent were tested to find the most appropriate drag model in this case. The volume fraction equation for secondary phase $p$ can be achieved from the continuity equations as shown in Eq.\,\eqref{eq:momentum}e.


\subsection{Kinetic theory of granular flow (KTGF)}
\begin{table}[hbt!]
	\centering
	\abovedisplayskip=0pt
	\belowdisplayskip=0pt
	\caption{Equations of the kinetic theory of granular flow.}\label{tab:ktgf}
	\begin{tabular}{ m{0.5cm} m{10.5cm} }\hline

	 & {\begin{subequations}\begin{flalign}
	 			& \mu_s  = \mu_{s,\mathrm{col}} + \mu_{s,\mathrm{kin}} + \mu_{s,\mathrm{fr}}&\\
	 			& \mu_{s,\mathrm{col}} = \frac{4}{5} \alpha_s \rho_s d_p g_{0,\mathrm{ss}} \left(1+e_{ss}\right) \left(\frac{\Theta_s}{\pi}\right)^{1/2}\alpha_s&\\
	 			& \mu_{s, k i n}=\frac{a_{s} d_{s} \rho_{s} \sqrt{\Theta_{s} \pi}}{6\left(3-e_{s s}\right)}\left[1+\frac{2}{5}\left(1+e_{ss}\right)\left(3 e_{ss}-1\right) \alpha_{s} \textsl{g}_{0,ss}\right]&\\
	 			& \mu_{s,\mathrm{fr}} = \frac{p_\mathrm{fr} \sin \varphi}{2 I_{2D}^{1/2}} &\\
				& \Theta_s = \frac{1}{3} \norm{\vec{v}_s^{\,\prime}}^2 &\\
				& 0 = \left(-p_s \overline{\overline{I}} + \overline{\overline{\tau_s}}\right):\nabla \vec{v_s} -\gamma_{\Theta_s} +\phi_{ls}&\\
				&\gamma_{\Theta_s}=\frac{12\left(1-e^2_{ss}\right) g_{0,ss}}{d_p \pi^{1/2}}\rho_s\alpha^2_s \Theta^{3/2}_s &\\
				& \phi_{ls}=-3K_{ls}\Theta_s &\\
					&p_{q}=\alpha_{q} \rho_{q} \Theta_{q}+\left(\sum_{p=1}^{N} \frac{d_{pq}^{3}}{d_{q}^{3}} p_{c, qp}\right) \rho_{q} \Theta_{q} &\\
					& \textsl g_{0,pq}=\frac{1}{\left(1-\alpha_{s}\right)}+\frac{3\left(\sum_{k=1}^{N} \frac{\alpha_{k}}{d_{k}}\right)}{\left(1-\alpha_{s}\right)^{2}\left(d_{p}+d_{q}\right)} d_{p} d_{q}&
				\end{flalign}\label{eq:ktgf}\end{subequations}}\\

		\hline\end{tabular}
\end{table}

The kinetic theory of granular flow was implemented to capture the effects of particle\textendash particle interactions on the flow behavior. The equations of frequently used KTGF sub\textendash models are given in Table\,\ref{tab:ktgf} in detail. The effective viscosity for the mixture with respect to the the granular viscosity is obtained by constitutive equations (Eq.\,\eqref{eq:ktgf}a) derived by \citet{syamlal1993mfix} and \citet{gidaspow1991hyd}. The collisional (Eq.\,\eqref{eq:ktgf}b \citep{gidaspow1991hyd}) and kinetic (Eq.\,\eqref{eq:ktgf}c \citep{syamlal1993mfix}) parts, and the optional frictional part (Eq.\,\eqref{eq:ktgf}d) are added to give the solids shear viscosity. In this way, the shear viscosity arising from particle momentum exchange due to translation and collision is taken into account. 

The granular temperature is expressed as Eq.\,\eqref{eq:ktgf}e. Mixture model solves the algebraic equation from the granular temperature transport equation as described by Eq.\,\eqref{eq:ktgf}f where a balance of energy is written. The terms on the right\textendash side of the equation represent the generation and the collisional dissipation of energy by the solid stress tensor, and the energy exchange between phases, respectively.

A general solids pressure formulation in the presence of more than one solid phases could be of the form of Eq.\,\eqref{eq:ktgf}i, where $d_{p q}=\frac{d_{p}+d_{q}}{2}$ is the average diameter, and $p_{c, qp}$ is the collisional part of the pressure between phases $q$ and $p$. In Eq.\,\eqref{eq:ktgf}i, $p_{c, q p}=2\left(1+e_{p q}\right) g_{0, p q} \alpha_{q} \alpha_{p} .$  When the number of solid phases is greater than 1, the radial distribution function between the $p^{th}$ and $q^{th}$ solid phases is expressed by \citet{syamlal1993mfix} as in Eq.\,\eqref{eq:ktgf}j. 

\subsection{Turbulence model}
In our previous study \citep{sadeghi}, we highlighted the suitability and performance of the Two\textendash Equation turbulence models ($k-\epsilon$ and $k-\omega$) for the multiphase slurry flows. Especially, based on the previous works in the literature and the theory of the turbulence models, we emphasized the acceptable and accurate prediction of the turbulent regieme in a multiphase flow using the $k-\omega$ model proposed by \citet{menter}. In the present study, at first both $k-\epsilon$ and $k-\omega$ models were examined, and the SST $k-\omega$ showed a better numerical stability leading to faster convergence, and excellent agreements between the CFD predictions and field data. 
\subsection{Rheological model}
\citet{rheology} studied the effect of particle size on the rheology of Athabasca oil sand slurries, and presented a comprehensive rheological study on the mixture of fine particles harvested from Mature Fine Tailing (MFT) and water which leads to formation of a non\textendash Newtonian fluid highly similar to the carrier fluid of this study. They performed the rheological measurements using three samples of fine particles with different size distributions named Fraction 1, Fraction 2, and Fraction 3 and various mixtures of them. Fraction 1 of their study has the closest size distribution to the fine particles in this study. Therefore, the Casson non\textendash Newtonian rheological model considered in this study through user\textendash defined function (UDF), which is a specific as carrier fluid viscosity:
\begin{equation} \label{eqn:Cassonmodel}
\tau^{1/2}=\tau_{\mathrm{y}}^{1/2}+\mu_{\mathrm{c}}^{1/2} \dot{\gamma}^{1/2}
\end{equation}
In Equation \ref{eqn:Cassonmodel}, $\tau$ is shear stress, $\tau_{y}$ represents yield stress, and $\mu_c$ is Casson viscosity \citep{rheology}. In case there is a gradual transition from Newtonian behavior to yield region, the Casson rheological model is appropriate to describe the carrier fluid viscosity 
	 \citep{casson}. The rheological behavior of tailings slurries has been successfully modeled using Casson rheological model by other researchers
	  \citep{britocasson,famcasson,ZHANGcasson}.

\subsection{Numerical methodology}
 Finite Volume Method (FVM) based commercial CFD software Fluent 2020R2 was used in this study and to solve the governing equations mentioned above. The velocity and volume fractions were specified at the inlet for all phases. At the wall, no\textendash slip boundary condition was chosen which applies to the mixture, and for the outlet zero gauge pressure was imposed. Computational model settings are summarized in the Table\,\ref{tab:CFDsettigns}. The pressure relaxation factor and momentum relaxation factor were set as 0.3 and 0.7. The volume fraction was set as 0.4, and the default values of other factors were used. Root mean square residuals were used, and the residuals for convergence were set as $10^{-4}$ and a fixed time step 0.01 utilized in this study.
 
 The flow conditions of the simulated cases such as the mixture velocity, solids fraction, carrier density and bitumen fraction can be seen in Table\,\ref{tab:expconditions}. These cases will be used for validation in section\,\ref{validationsection}, and Case 2 will be used as the base case for grid\textendash independence check and further investigations.

\begin{table}[!ht]
\caption{List of different models and schemes used in multiphase mixture model for modeling three\textendash phase (liquid\textendash solid\textendash liquid) slurry flow.}
\vspace{0.3cm}
\label{tab:CFDsettigns}
\resizebox{\textwidth}{!}{%
\begin{tabular}{ll}
\specialrule{.1em}{.05em}{.05em} 
\multicolumn{1}{c}{Model}                      & \multicolumn{1}{c}{Scheme} \\ \specialrule{.1em}{.05em}{.05em} 
Multiphase model                                 & Mixture           \\
Viscous model                                    & $k$\textendash $\omega$ SST model              \\
Volume Fraction Parameters                       & Implicit Scheme             \\
Pressure\textendash velocity coupling                       & Phase Coupled SIMPLE Scheme \\
Spatial discretization\textendash Gradient                  & Least Squares Cell Based    \\
Spatial discretization\textendash Pressure                  & PRESTO    \\
Spatial discretization\textendash Momentum                  & First Order Upwind         \\
Spatial discretization\textendash Volume Fraction           & First Order Upwind       \\
Spatial discretization\textendash Turbulent Kinetic Energy  & First Order Upwind         \\
Spatial discretization\textendash Specific Dissipation Rate & First Order Upwind         \\
Transient Formulation                            & First Order Implicit         \\

Granular viscosity                               & \citet{syamlal1989computer}              \\
Solid pressure                                   & \citet{lun1984kinetic}                  \\
Granular temperature                             & Algebraic                  \\
Frictional Pressure (pascal)                     & Based\textendash KTGF             \\
Radial distribution                             & \citet{syamlal1989computer}           \\
Drag model: Liquid\textendash solid                             &   \citet{gidaspow1991hyd}   \\
Drag model: Liquid\textendash Liquid                            &   Symmetric \citep{fluent2011ansys} \\
Slip velocity                            &   \citet{manninen} \\

\specialrule{.1em}{.05em}{.05em} 
\end{tabular}%
}
\end{table}

 \begin{table}[hbt!]
 \caption{Flow conditions of the simulated cases.}
    \centering
    \begin{tabular}{P{1cm}|P{1.5cm}|P{3cm}|P{2cm}|P{3cm}}
    \hline
  Case & $\alpha_v$ & $\rho_{carrier}$\,($kg/m^3$) & $v_m$\,(m/s) & Bitumen volume fraction \\ \hline
   (1) & 0.238 & 1335 & 5.62 & 0.0025  \\ \hline
   (2) & 0.23 & 1278 & 5.35 & 0.0033  \\ \hline
   (3) & 0.237 & 1328 & 5.43 & 0.0029 \\ \hline
   (4) & 0.269 & 1227 & 5.54 & 0.003 \\ \hline
   (5) & 0.26 & 1221 & 5.20 & 0.0037 \\ \hline
   (6) & 0.264 & 1196 & 5.21 & 0.0032 \\ \hline
   (7) & 0.223 & 1222 & 5.76 & 0.003 \\ \hline
   (8) & 0.222 & 1220 & 5.18 & 0.0028 \\ \hline
   (9) & 0.273 & 1254 & 5.30 & 0.0044 \\ \hline
   (10) & 0.238 & 1332 & 5.64 & 0.003 \\ \hline
   
   \end{tabular}
    \label{tab:expconditions}
\end{table}



\section{Results and discussion}

\subsection{Composition of tailings slurry}

\begin{figure}[hbt!]
\centering
  
  \includegraphics[width=0.99\textwidth]{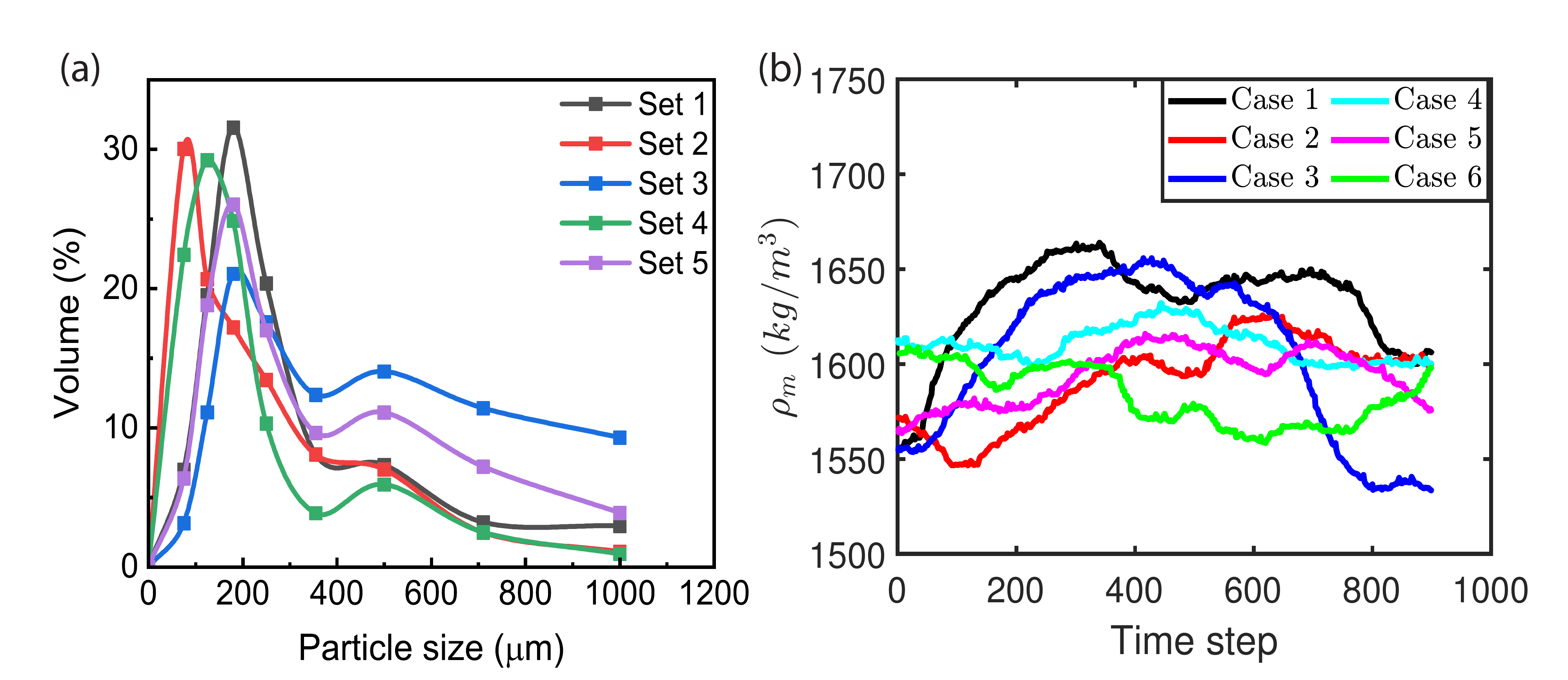}
\caption{[Color] Plots of (a) PSD of the coarse particles for three measurements and (b) mixture density for six representative cases.}\label{fig:psd-density}

\end{figure}

The PSD information in five different measurements is presented in Fig.\,\ref{fig:psd-density}a. As discussed earlier, a sieving method is used to determine the size distribution for solid particles, with mesh sizes from 1.3 to 2000\,$\mu m$. Particles with sizes below 44\,$\mu m$ are considered as fines and incorporated in the carrier fluid, and the fraction for particles larger than 2000\,$\mu m$ is zero in all of the cases. With these considerations, eight bins for coarse solid particles are formed, from $d_p$ = 75 to 1000\,$\mu m$. The compositions data is reported in the form of mass fraction, which is first converted to volume fraction with having the densities of solids and water, and later on, used to calculate the carrier fluid density. It must be noted that Set 1 in Fig.\,\ref{fig:psd-density} is the PSD information for the base case and used for all of the cases simulated in the validation section. The slurry density of three representative time windows is depicted in Fig.\,\ref{fig:psd-density}b. 

\subsection{Velocity profiles and flow conditions}
\label{section:expresults}
\begin{figure*}[ht!]
\centering
\centering
 
  \includegraphics[width=0.99\textwidth]{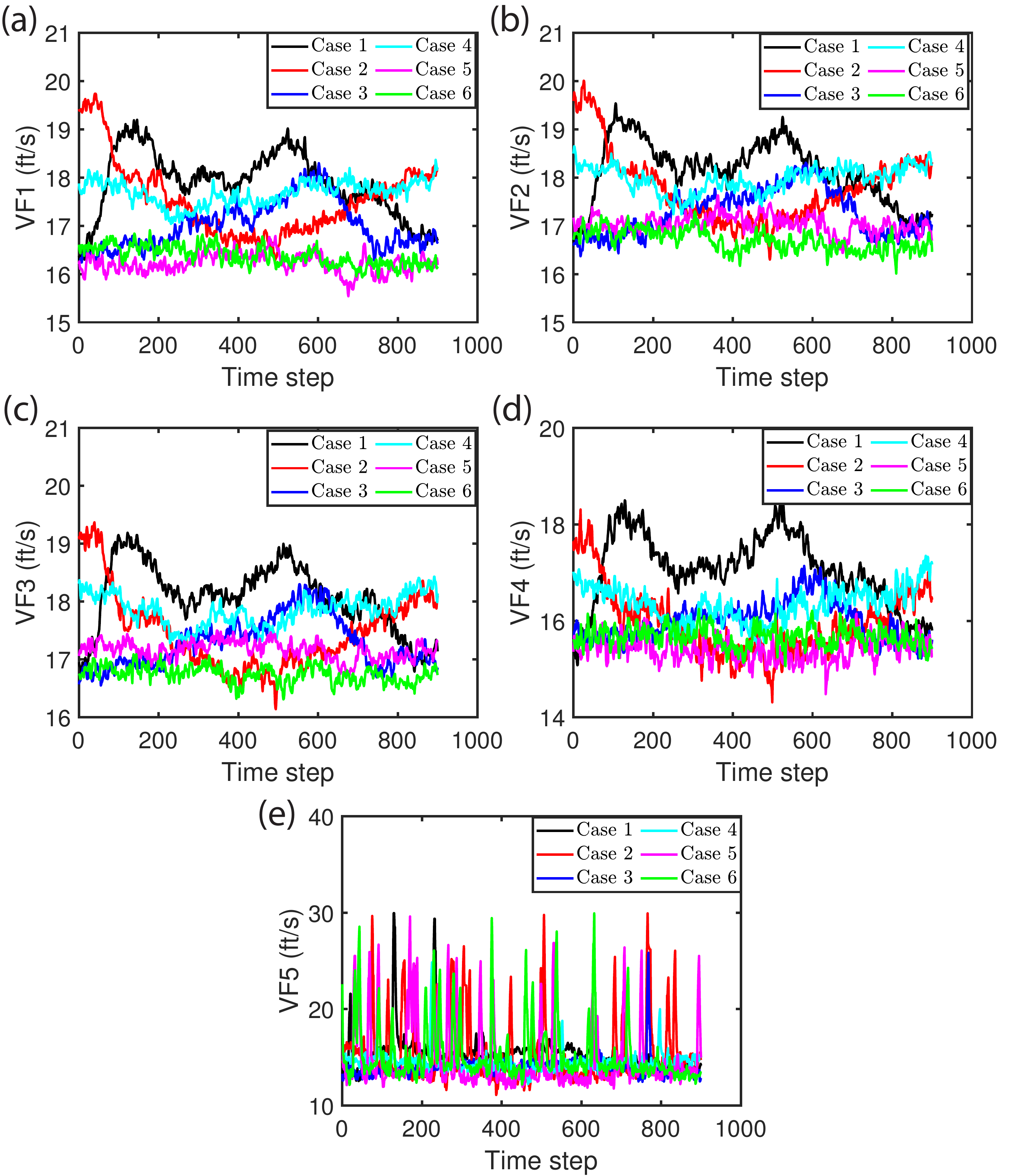}
\caption{[Color] Plots of (a) VF1, (b) VF2, (c) VF3, (d) VF4, and (e) VF5 vs time step for six representative cases.}\label{fig:Fielddata-velocity}
\end{figure*}

Fig.\,\ref{fig:Fielddata-velocity} represents the values of the velocities over 900 time steps for six representative cases (Case 1 to Case 6 in Table\,\ref{tab:expconditions}). The velocity values are fairly consistent with a narrow range of fluctuations for VF1 to VF4. However, as observable in Fig.\,\ref{fig:Fielddata-velocity}e, VF5 is prone to more frequent and significant fluctuations. This is observed almost for the entire set of data points, indicating that the velocity values at the pipe invert are significantly influenced by some sort of interference. As mentioned, the velocity values are calculated based on the force exerted by the turbulence eddies on the pipe wall, and due to the presence of solid particles and frequent collisions between them and the pipe wall, the force sensed by the device is not only because of the eddies and contains the effect of solid particles. To eliminate this effect from the average value of VF5, the significantly large and small values compared to the normal range (VF5 $>18\,ft/s$ and VF5 $< 14\,ft/s$) were removed from the range used for averaging.

\begin{figure*}[ht!]
\centering
\centering
 
\includegraphics[width=0.99\textwidth]{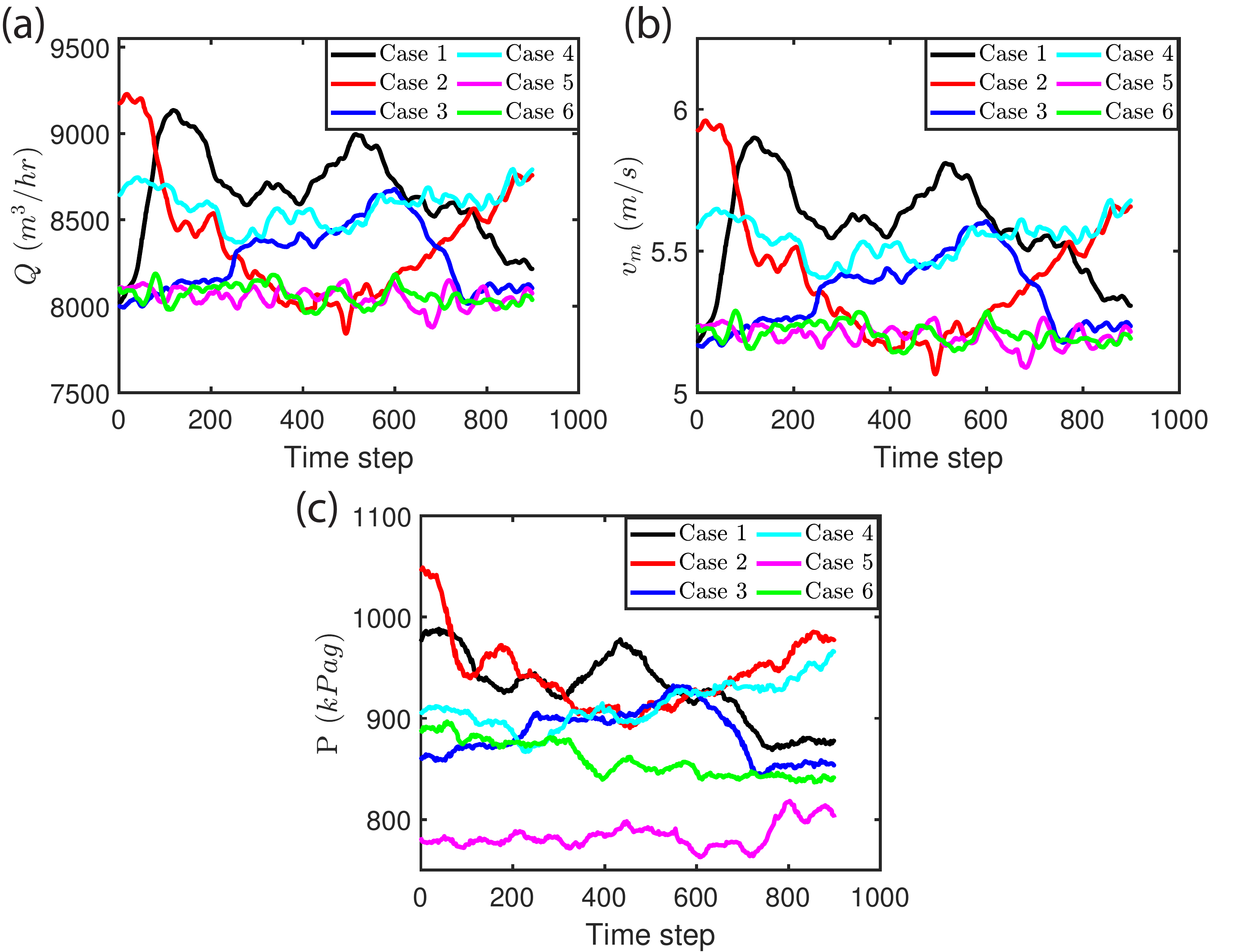}
\caption{[Color] Plots of (a) volumetric flowrate, (b) mixture velocity, and (c) gauge pressure at the first pump discharge vs time step for six representative cases.}\label{fig:Fielddata1}
\end{figure*}

The flow characteristics for the representative cases are shown in Fig.\,\ref{fig:Fielddata1} in which panels a to c depict the flowrate, mixture velocity, and pressure at the discharge of the first stage, respectively. The mixture velocity is calculated with respect to the flowrate, which is the reason for the similar trends seen in panels (a) and (b). In each time window, there are specific ranges where the fluctuations are weaker and the flow parameters are more consistent.

Over around a million data points collected, the selection of a proper time window must be made carefully, with monitoring the fluctuations and significant shifts in the flow conditions. A 30\textendash minute time window is selected as the initial step, and the data points are plotted against the time steps. If the oscillations over the entire time window or a portion of it are insignificant and the flow conditions are relatively consistent, that time window is chosen for simulations. Averaging the values over the selected time window is performed, and the cases are prepared.

\subsection{Grid independence}

\begin{figure}[hbt!]

\centering
\includegraphics[width=0.85\textwidth]{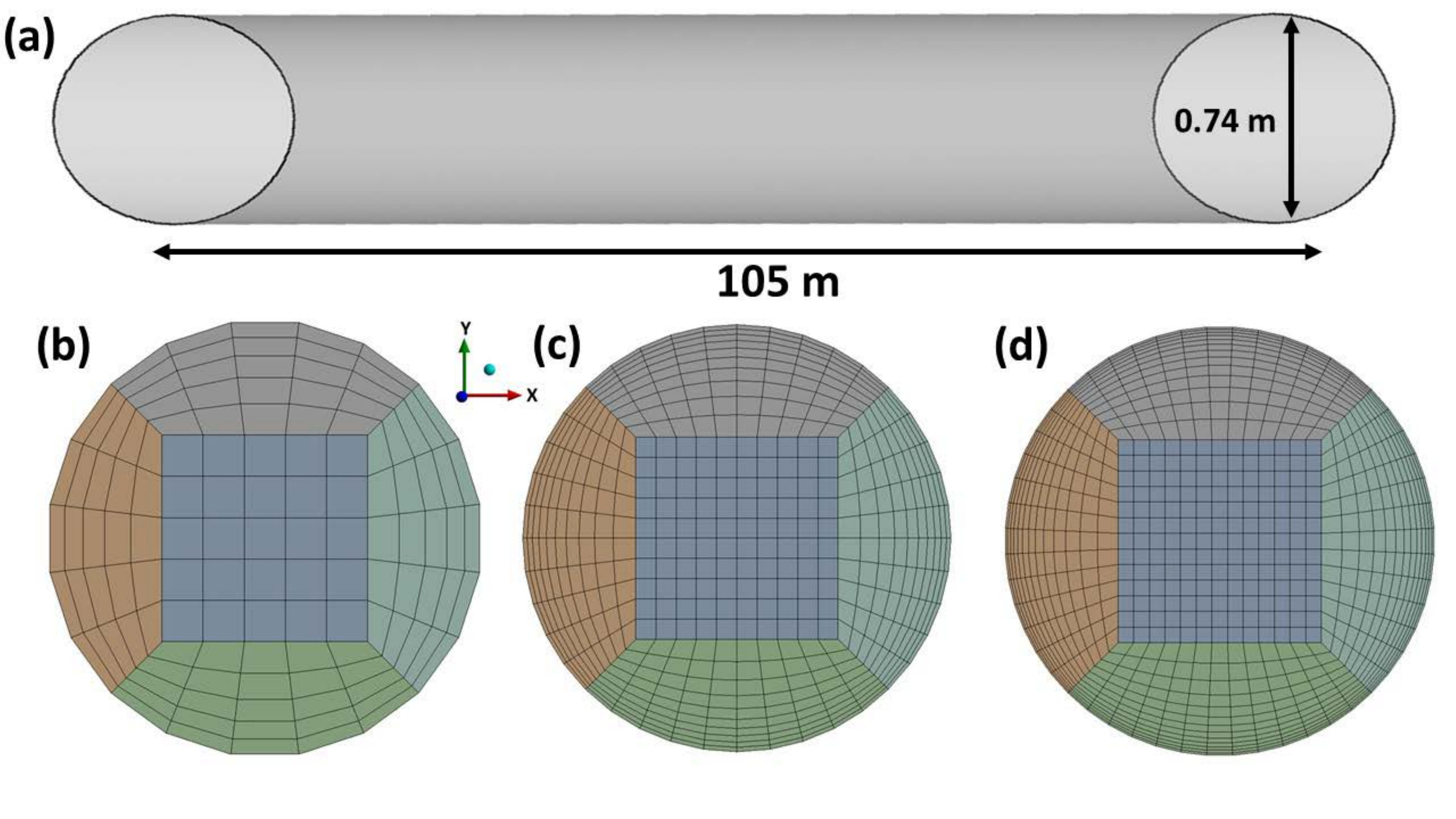}

\caption{ (a) Schematic of the three\textendash dimension computational geometry with the channel dimensions. Axial view of different grid structure (b) coarse, (c) fine, and (d) finer meshes.} \label{fig:gridstructure}
\end{figure}

\begin{figure*}[ht!]
\centering
\centering
 
 \includegraphics[width=1\textwidth]{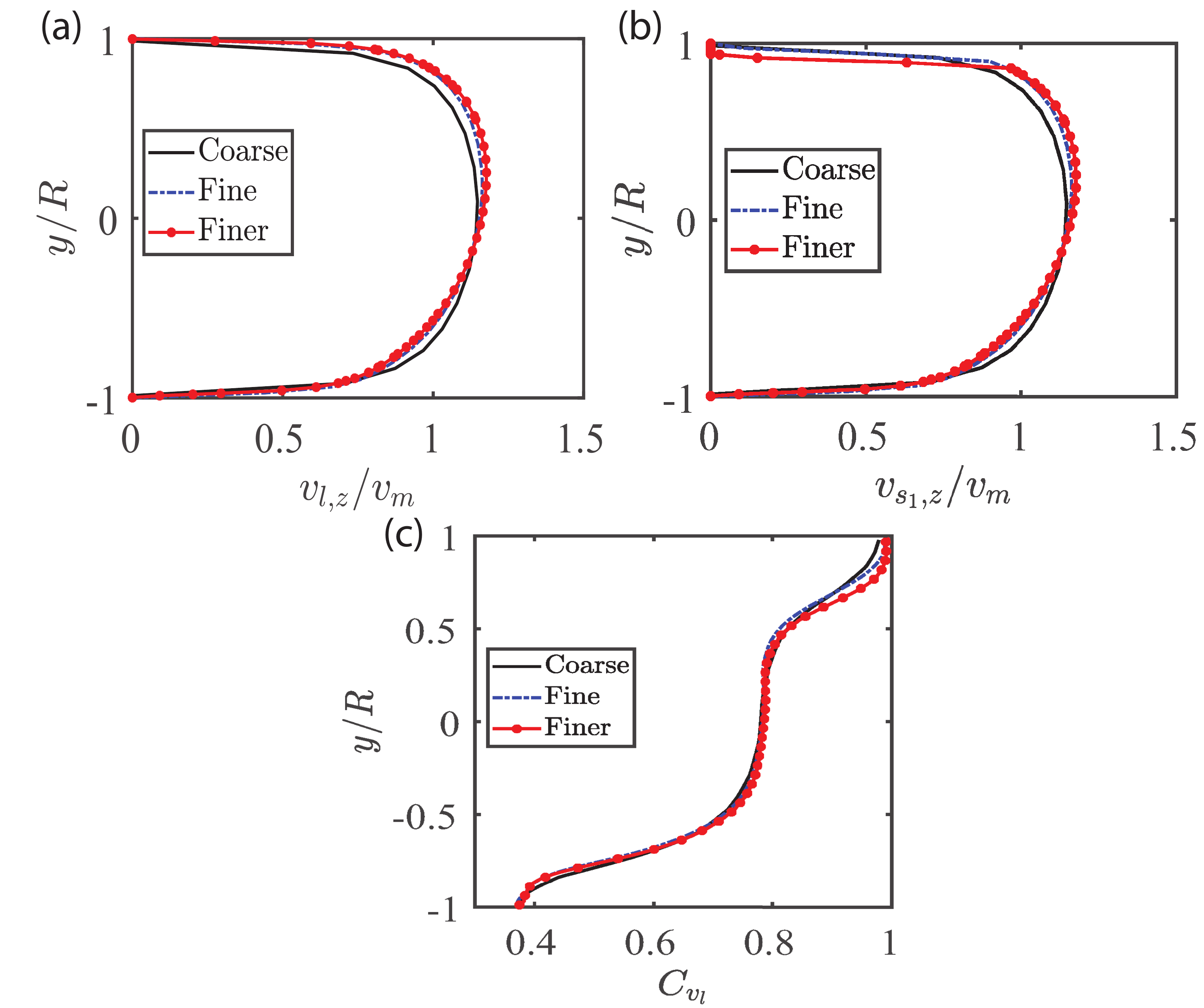}
\caption{Plots of (a) carrier velocity, (b) first solid phase velocity, and (c) carrier volume fraction distributions with the coarse, fine and finer grids. (\begin{math}D = 74.0\,\mathrm{cm}, \alpha_v = 0.23, v_m = 5.35\,m/s, \rho_s = 2650\,\mathrm{kg/m^3},  \rho_l = 1278\,\mathrm{kg/m^3},
\tau_y = 0.04696\, \mathrm{Pa}, \mu_c = 0.00165\,\mathrm{Pa^{1/2}.s^{1/2}}) \end{math}}\label{fig:GridIndependence}
\end{figure*}

\begin{figure}[h!]
\centering
  \includegraphics[width=1\textwidth]{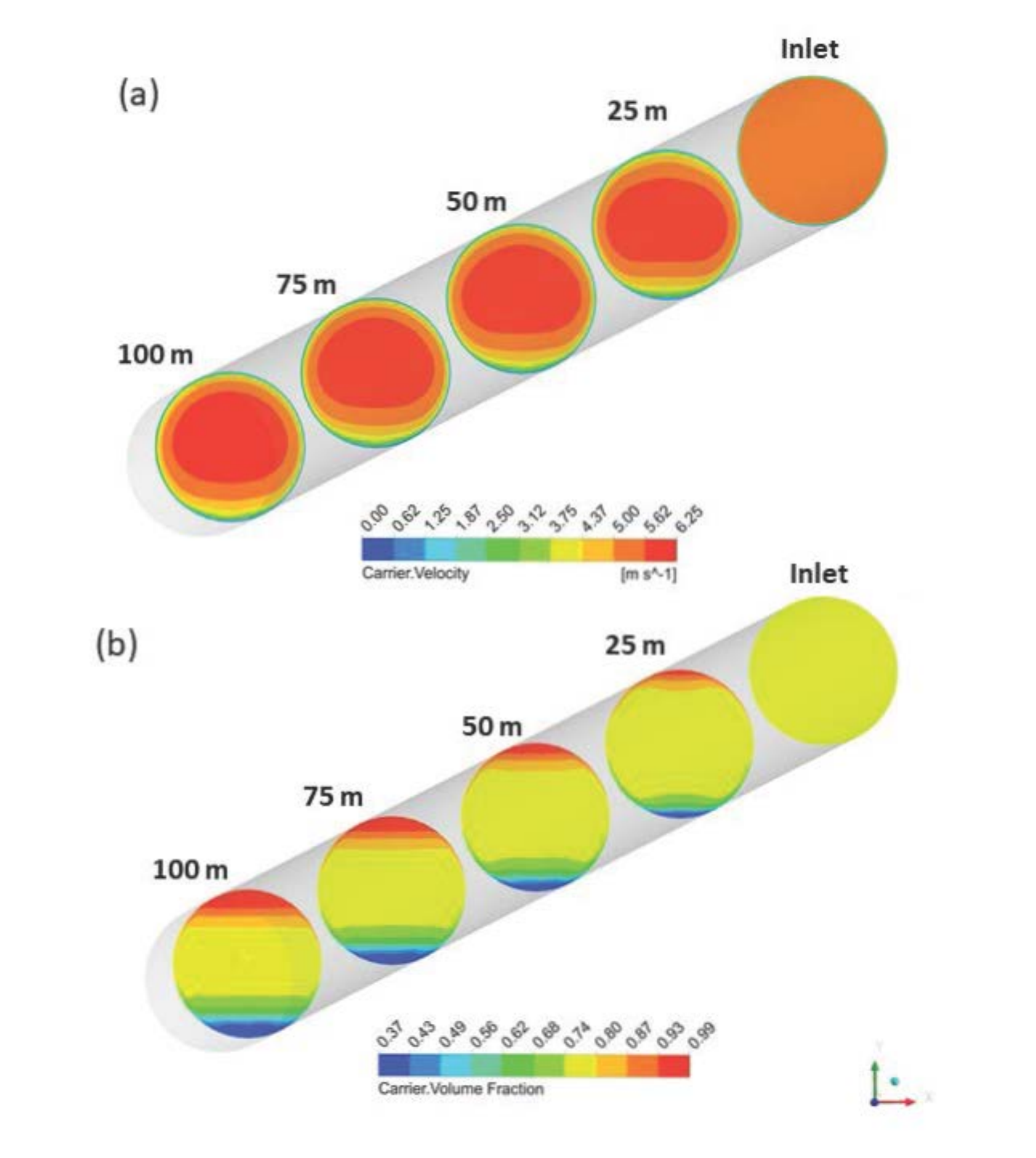}\\
\caption{[Color] Contours of (a) carrier velocity and (b) volume fraction distributions at the inlet and axial distances of 25, 50, 75, and 100 m. (\begin{math}D = 74.0\,\mathrm{cm}, \alpha_v = 0.23, v_m = 5.35\,m/s, \rho_s = 2650\,\mathrm{kg/m^3},  \rho_l = 1278\,\mathrm{kg/m^3},
\tau_y = 0.04696\, \mathrm{Pa}, \mu_c = 0.00165\,\mathrm{Pa^{1/2}.s^{1/2}}) \end{math}}\label{fig:fullydev}
\end{figure}

Three grids with similar structure but different number of elements (109125, 438000, and 741065 elements) were designed as shown in Fig\,\ref{fig:gridstructure}\,b\textendash d. The details of the flow conditions of the simulated case can be seen in Table \ref{tab:expconditions} as Case 2. Fig.\,\ref{fig:GridIndependence}a\textendash c exhibit the comparison of the calculated carrier velocity distribution, the first solid phase velocity distribution and carrier concentration distribution, respectively at a distance of 100\,m from the inlet over the three designed grids. The distributions are almost uniform for the coarse, fine, and finer grids. Especially for the fine and finer meshes, nearly no differences are seen for both carrier and first solid velocity and concentration distribution, proving that the so\textendash called 'fine' mesh in this section is appropriate to be used for all of the simulations, as the results are independent of the number of elements. Fig.\,\ref{fig:fullydev} shows the contours of velocity and volume fraction of the carrier fluid on separate cross\textendash sectional planes at the inlet, and axial distances of 25, 50, 75, and 100\,m. As depicted in Fig.\,\ref{fig:fullydev}, the distributions do not change after 50\,m of the pipe inlet for both velocity and volume fraction. This analysis demonstrates that the flow has reached the fully\textendash developed state before the pipe outlet, where the flow behavior is independent of the axial distance. Accordingly, for the analysis of the CFD results in the the radial direction, a plane at z = 100\,m was used to export the data.

\subsection{Model validation}
\label{validationsection}

 To demonstrate the accuracy and validity of the developed model, CFD simulations were compared with field data of velocity distribution and pressure gradient. The accordance of the CFD predictions with field data demonstrates that the physics of the system has been well defined, and the model can be considered a reliable tool to perform a parametric study on a similar system. The comparisons between CFD\textendash predicted and measured velocity distribution and pressure gradient are presented in this section.
\begin{figure}[hbt!]

\centering
\includegraphics[width=0.99\textwidth]{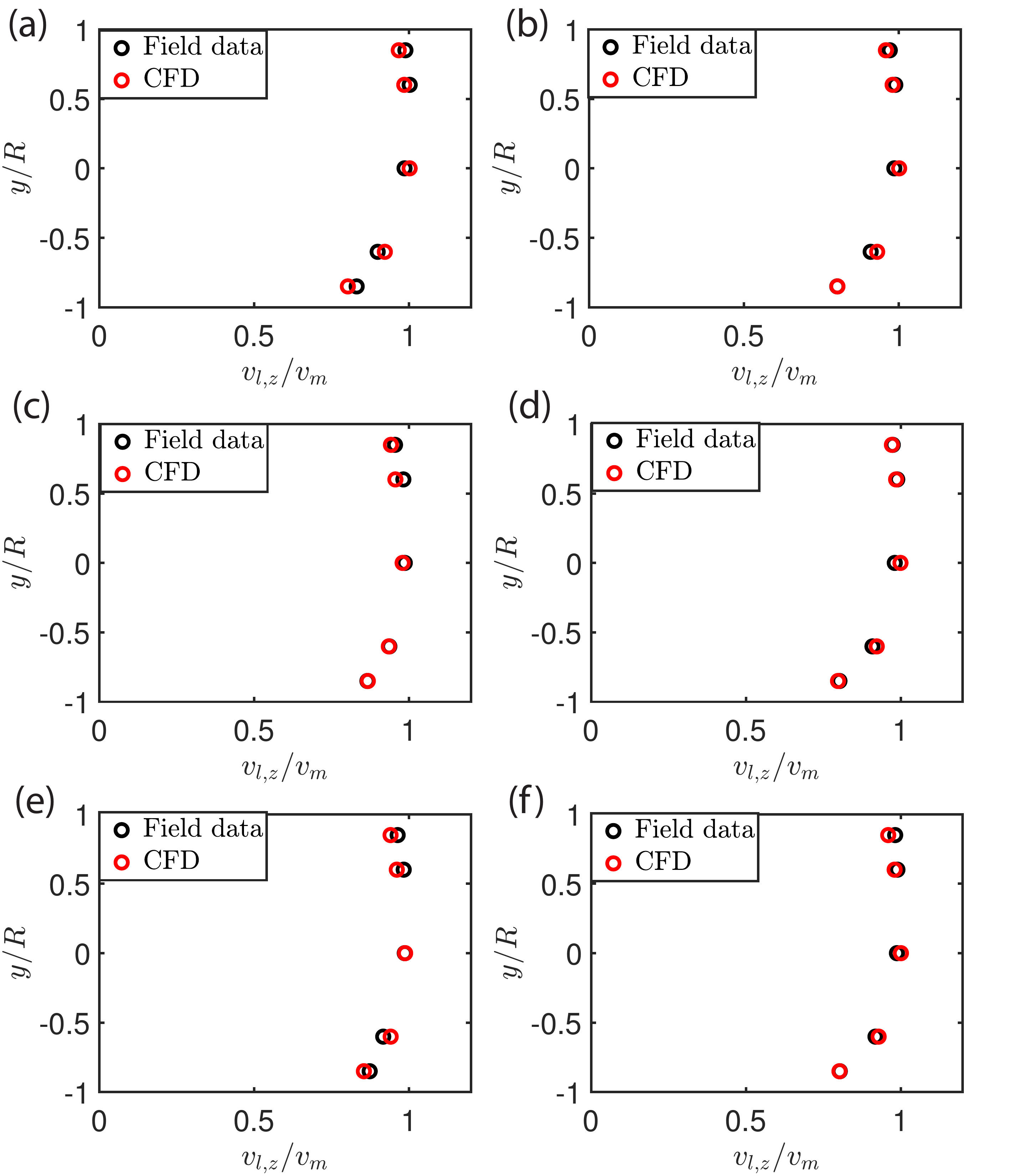}
\caption{[Color] Comparisons of the velocity distributions predicted by CFD (the red circles which represent the interpolated CFD values) and measured at the field (black circles) for six different cases (a) Case 2, (b) Case 3, (c) Case 7, (d) Case 8, (e) Case 9, and (f) Case 10.  }\label{fig:validationvel}
\end{figure}

The details of the flow conditions for the cases simulated for the validation purpose are provided in Table \ref{tab:expconditions}. Fig.\,\ref{fig:validationvel} depicts the comparison between the CFD\textendash predicted velocity distribution and the measured ones at the field for six different cases. The results demonstrate exceptional agreement between the carrier fluid velocity data exported from the CFD simulations and measured at the field. For most of the data points, the CFD and field values overlap, with insignificant discrepancies for some data points. 


\begin{figure}[hbt!]
\centering

  \includegraphics[width=0.7\textwidth]{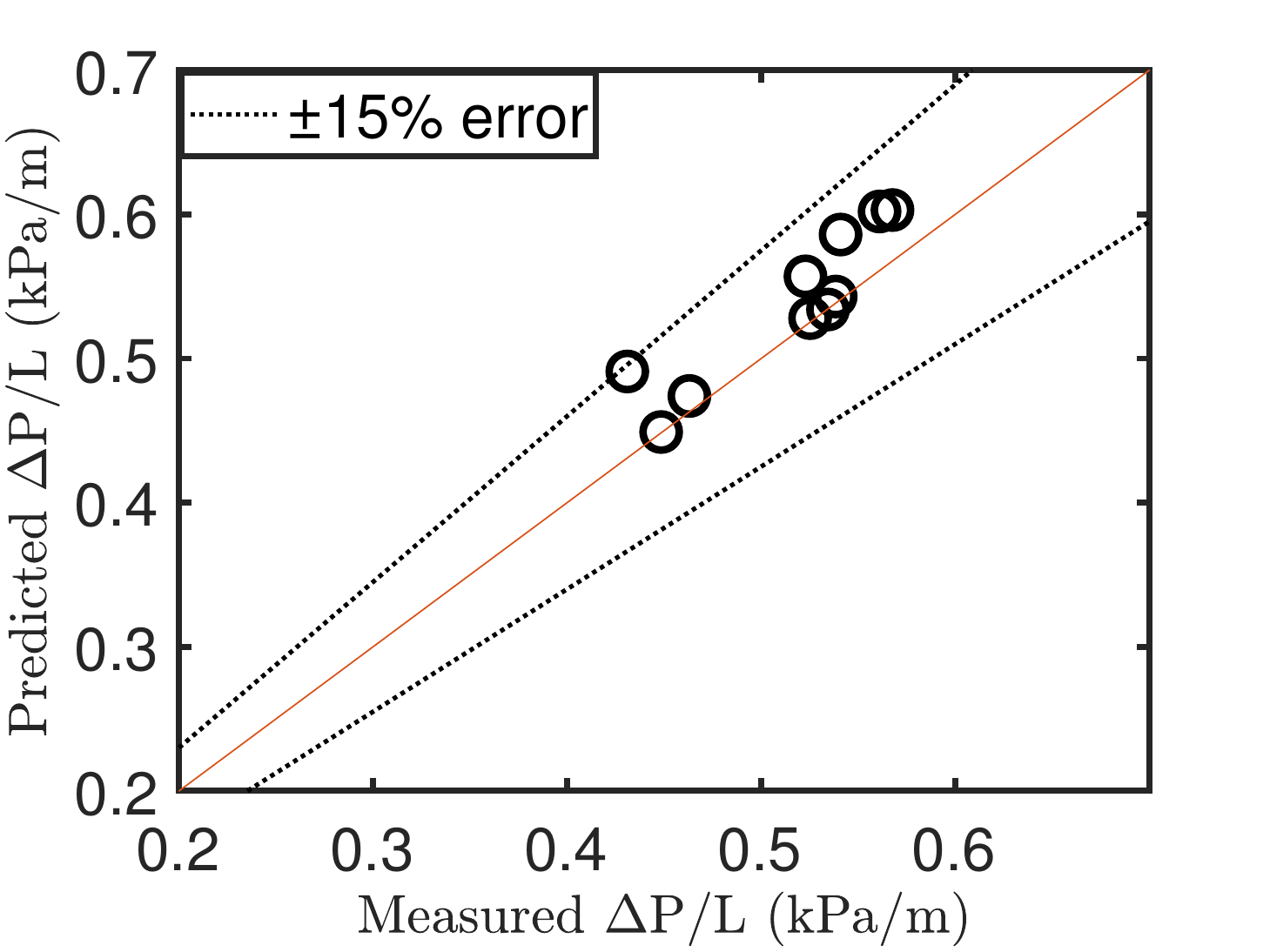}\\
\caption{Parity plot of frictional pressure drop obtained from CFD vs.~field data. The solid line is the bisector and the two dashed lines denote errors of $\pm15$\%.}\label{fig:delp_valid}
\end{figure}

The accuracy of the model and validity of the physics defined for the system was also examined by the comparison of pressure gradients predicted by CFD and measured pressure at the industrial pipeline. Overall, the comparison was made for 10 distinct cases over 10 different time windows, and the results are presented in Fig.\,\ref{fig:delp_valid}. For all of the cases, the predictions are in accord with the field data with an error margin of less than 15\% which can be considered highly accurate for an industrial\textendash scale pipeline. The details of the flow conditions, such as the mixture velocity, carrier density, total solids concentration, etc., are presented in Table \ref{tab:expconditions}. The frictional pressure drop inside a pipe results from particle\textendash wall and particle\textendash particle interactions \citep{wilson2006slurry}. The field measurements and CFD predictions show an increase in the frictional pressure gradient as the mixture velocity and total solids concentrations increase. A higher mixture velocity and solids concentration lead to enhanced and more frequent collisions between particles and with the pipe wall, resulting in a higher pressure drop. 

In summary, the model was proven to be accurate and sufficiently sophisticated in physics based on the comparison of CFD\textendash predicted velocity distribution and pressure gradient with field measurements.

\subsection{Sensitivity analysis}

\subsubsection{Selection of the drag model}
\begin{figure}[!ht]

\centering
\centering
\includegraphics[width=14cm, height=8cm]{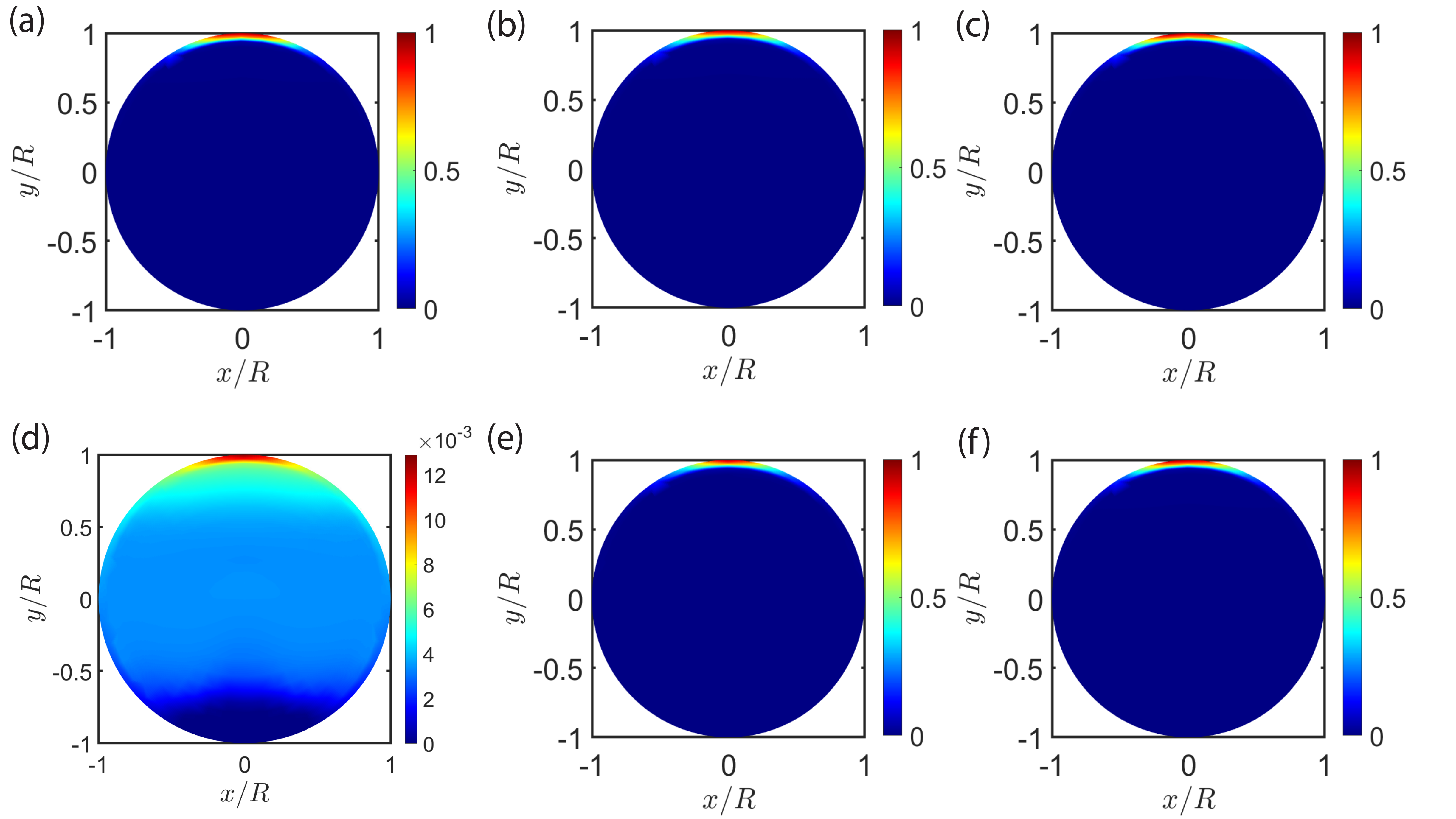}
\caption{[Color] Contours of concentration distribution of bitumen with using different drag models between the carrier fluid and bitumen. The drag models are: (a) schiller\textendash naumann, (b) morsi\textendash alexander, (c) universal\textendash drag, (d) symmetric, (e) ishii\textendash zuber, and (f) tomiyama }\label{fig:dragmodels}
\end{figure}

Choosing an appropriate drag model is crucial in the CFD simulation of any multiphase flow with mixture model as it will directly influence the slip velocity which is the only parameter that distinguishes the phases in mixture model. In Ansys Fluent, the available drag models are (a) schiller\textendash naumann \citep{schiller}, (b) morsi\textendash alexander \citep{morsi_alexander_1972}, (c) universal\textendash drag \citep{kolev-universal}, (d) symmetric \citep{fluent2011ansys}, (e) ishii\textendash zuber \citep{ishii}, and (f) tomiyama \citep{tomiyama}. Fig.\,\ref{fig:dragmodels} shows the contours of bitumen concentration distribution in radial direction with different drag models between bitumen and the carrier fluid. The results show that except for the symmetric drag model, all other drag models lead to total accumulation of bitumen droplets at the pipe obvert. Data collection from precise location for bitumen concentration distribution is challenging. We compared the bitumen concentration in two samples collected from the top and middle parts of the pipe, and bitumen concentration at the top region was twice as high as in the core region.  On this basis, the symmetric model was selected as the carrier\textendash bitumen drag model for this study.

\subsubsection{Bitumen droplet size}
\begin{figure}[!ht]

\centering
\centering
\includegraphics[width=0.99\textwidth]{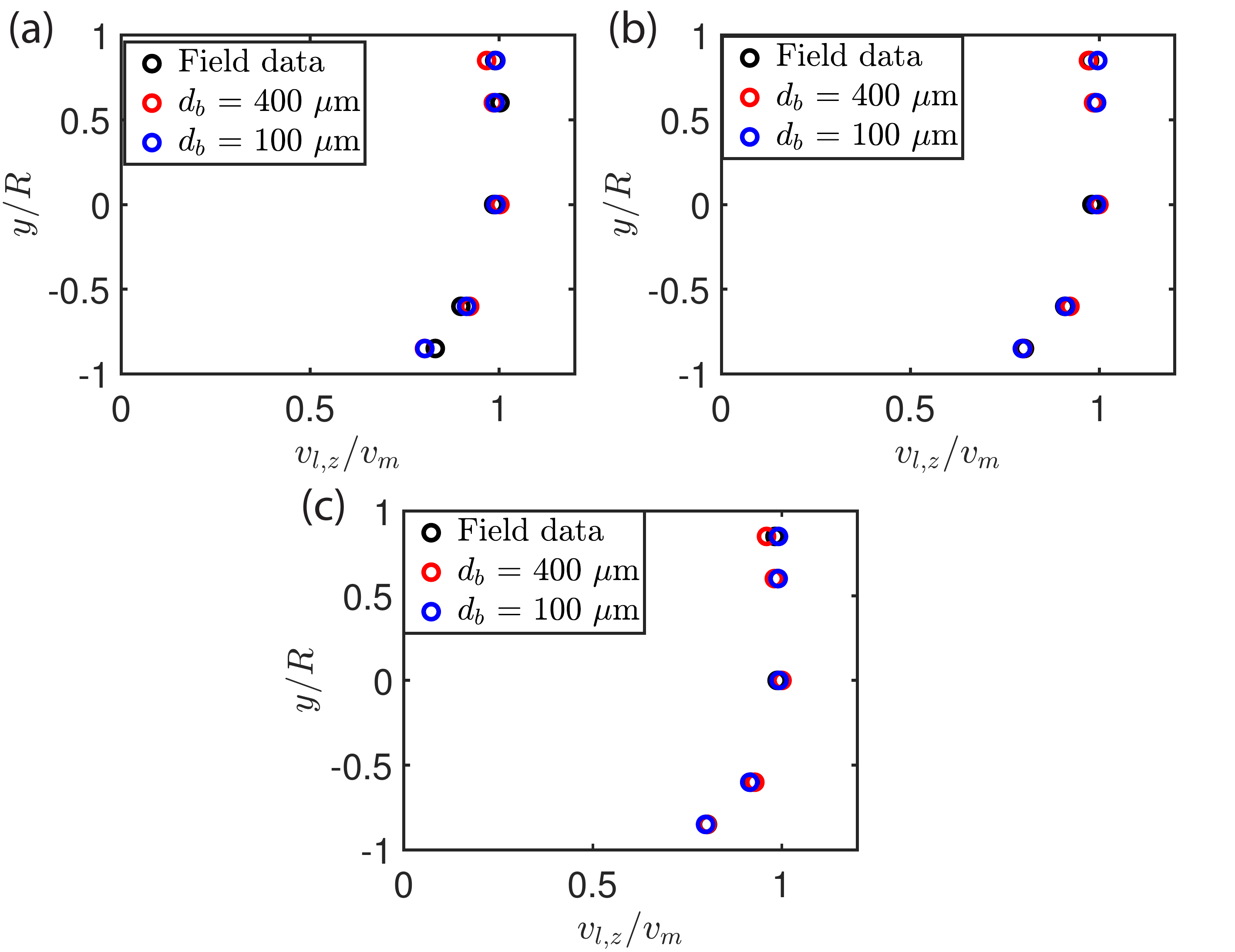}

\caption{[Color] Effect of bitumen droplet size on the velocity distribution for three different cases (a) Case 2, (b) Case 7, (c) Case 10.}\label{fig:bitsize}
\end{figure}

As bitumen is a dispersed phase in the form of droplet, a diameter needs to be considered for these droplets in the simulation. There are multiple studies in the literature for size characterization of bitumen droplets and aggregates in the flotation column, however, no data has been reported for bitumen droplet size in the flow in a horizontal pipeline. For the size distribution of bitumen droplets in a flotation column, it has been reported that most droplets are less than 800 $\mu$m in size \citep{MALYSA-size1,MALYSA-size2}. On this basis, a comprehensive analysis was performed to find the appropriate diameter for bitumen droplets in which the CFD prediction are in accord with the field data with the lowest error. The maximum diameter considered for the droplet was 800\,$\mu$m as it was assumed that the droplets can not be larger than the aggregates in the flotation column, and the lower band was considered to be 100\,$\mu$m. It was found that the flow behavior is almost insensitive to the bitumen droplet size. This was expected as the volume fraction of bitumen is low and the considered range for the size is not very broad. Fig.\,\ref{fig:bitsize} shows the velocity distribution is in the same form with droplets sizes of 100\,$\mu$m and 400\,$\mu$m for three representative cases. However, the diameter of 400\,$\mu$m overall gave better predictions of velocity and pressure gradient, and was selected as the input for the simulations in this study.

\subsection{Bitumen concentration distribution}

\begin{figure}[hbt!]
\centering

  \includegraphics[width=1\textwidth]{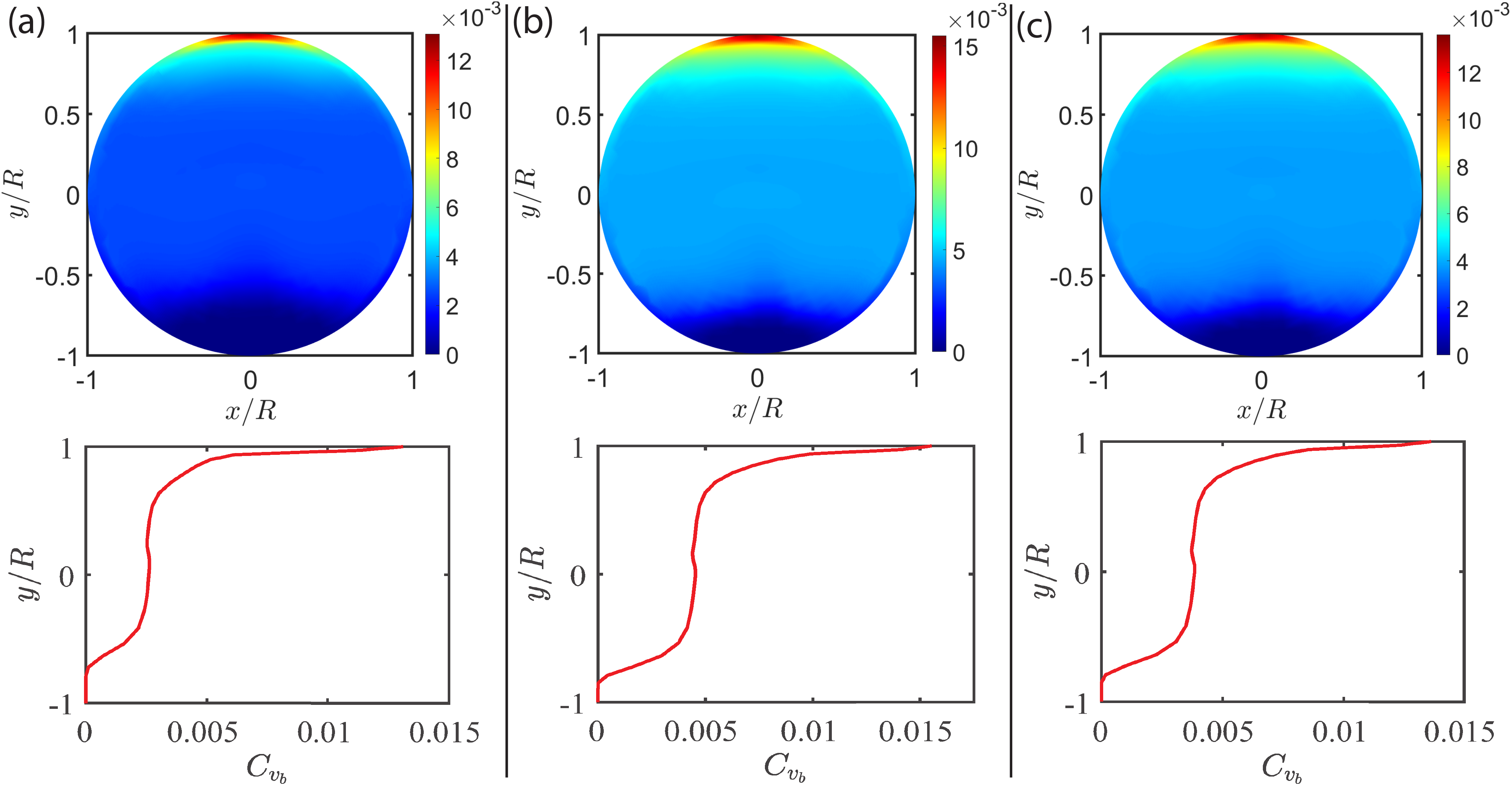}\\
\caption{[Color] Contours of bitumen concentration distribution and plots of bitumen concentration distribution on the vertical centerline for three representative cases (a) Case 1, (b) Case 9, and (c) Case 5.}\label{fig:bitdistribution}
\end{figure}

Fig.\,\ref{fig:bitdistribution} depicts the contours of bitumen concentration distribution in radial direction and plots of bitumen concentration on the vertical centerline at a distance of 100\,m from the inlet for three representative cases. As shown in Fig.\,\ref{fig:bitdistribution}, bitumen droplets mostly move to the top region of the pipe due to a lower density, but some droplets are still dispersed into the flow and bitumen concentration in the core is around half of the top region of the pipe. The distribution of bitumen droplets in the radial direction is a result of competition between the buoyancy effect which leads to accumulation of the drops at the top part, and turbulent dispersion which tries to bring the droplets back into the flow. This prediction of the trend in bitumen concentration distribution is crucial for future analysis on separation of the bitumen or any other third phase as long as it is a fluid and immiscible with water.

In the first section of Table\,\ref{tab:bitdistribution}, the distribution of bitumen droplets at the top one\textendash fourth (upper 25\,\%) and top half (upper 50\,\%) of the pipe (cross\textendash sectional view) is presented for the cases shown in Fig.\,\ref{fig:bitdistribution}. This analysis helps to design an efficient method for further treatment of the slurry for bitumen separation, with targeting the appropriate section of the pipe where a high percentage of droplets reside. As the results demonstrate, the majority of bitumen droplets are at the top half of the pipe, with around 72\,\% for Case 1 and 67\,\% for the other two cases. Therefore, the treatment of the flow of the top half section of the pipe can lead to an acceptable recovery of bitumen.

\begin{table}[hbt!]
\centering
\begin{tabular}{|P{4cm}|P{2.5cm}|P{2.5cm}|}

\hline
Case         & Upper 25\% & Upper 50\%   \\ \hline
1         &      47.02      &     71.97        \\ 
5        &           42.31       &     67.37      \\ 
9         &        41.27    &      66.27      \\ \hline
$v_m$ = 3.5 m/s  &      48.96      &     74.09         \\ 
$v_m$ = 4.5 m/s    &      45.21      &      70.77      \\ 
$v_m$ = 5.5 m/s   &        42.93    &          68.58    \\ 
$v_m$ = 6.5 m/s   &      41.13      &       66.68       \\ \hline
PSD: Set 1    &   43.08 &  68.65         \\ 
PSD: Set 2          &         43.7   &     73.16        \\ 
PSD: Set 3         &   43.35 &         71.48      \\ 
PSD: Set 4       &      43.7      &     75.28         \\ \hline
$\mu_c$ = 0.001 $Pa^{1/2}.s^{1/2}$ &       44.26  &        70.8      \\ 
$\mu_c$ = 0.003 $Pa^{1/2}.s^{1/2}$ &        42.52    &          67.57    \\ 
$\mu_c$ = 0.006 $Pa^{1/2}.s^{1/2}$ &        41.35    &           65.52    \\ 
$\mu_c$ = 0.01 $Pa^{1/2}.s^{1/2}$  &         40.31   &       63.87      \\ \hline
$\theta$ = -10                    &       43.19     &        68.12       \\ 
$\theta$ = -5                    &      44.15      &       69.12       \\ 
$\theta$ = 5                    &      44.0      &       70.29       \\ 
$\theta$ = 10                  &     42.53       &     69.19       \\ \hline

\end{tabular}
    \caption{Bitumen distribution at top 25\% and 50\% sections of the pipe. Note that the mentioned values are the ratio of bitumen volume fraction at each section to the total bitumen volume fraction in percentage. Case 2 is the base case for parametric study. }
    \label{tab:bitdistribution}
\end{table}

\subsection{Solids concentration distribution}

\begin{figure}

\centering
\includegraphics[width=1\textwidth]{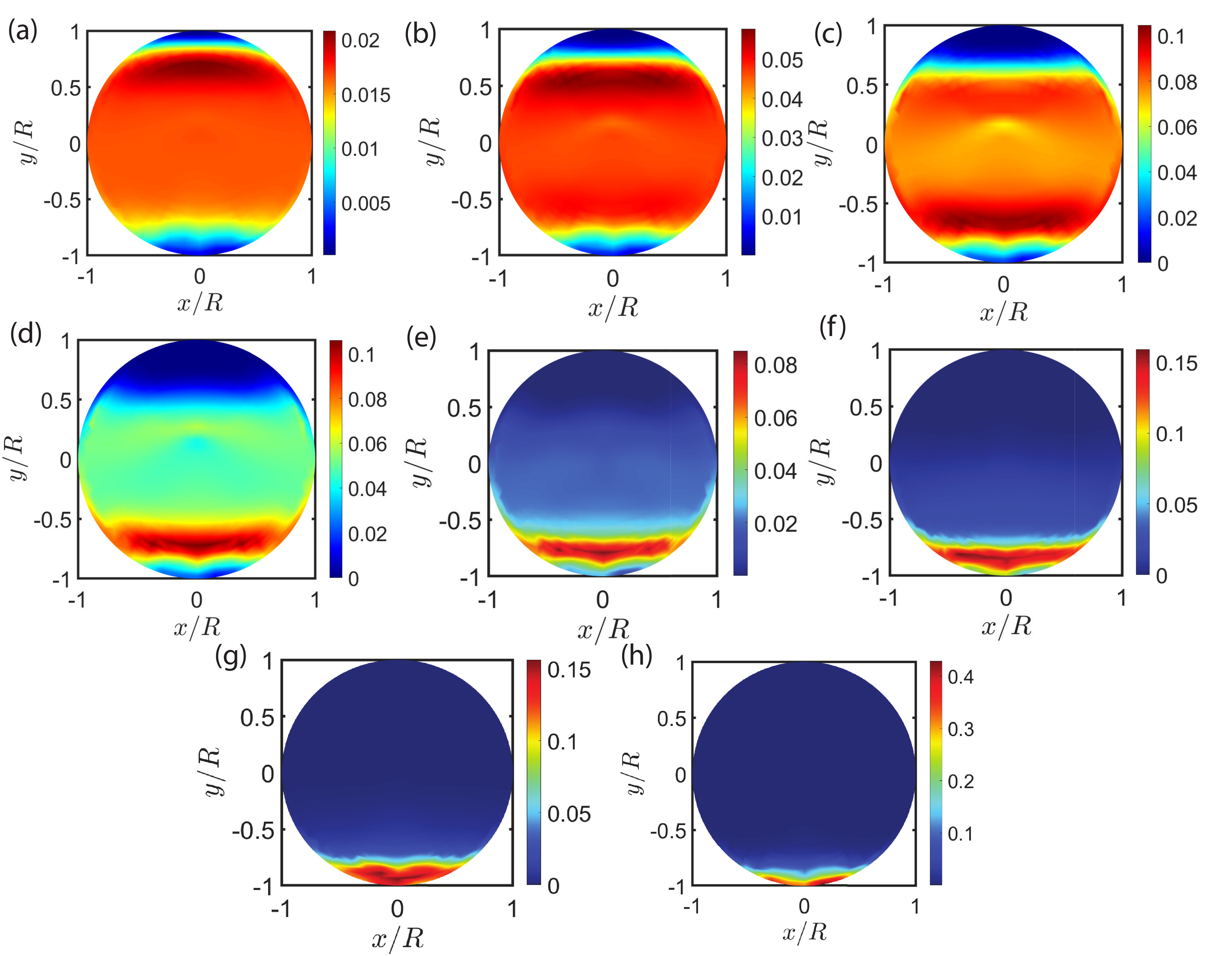}
\caption{[Color] Contours of concentration distribution of solid particles with sizes ($d_p$) of (a) 75\,$\mu$m, (b) 125\,$\mu$m, (c) 180\,$\mu$m, (d) 250\,$\mu$m, (e) 355\,$\mu$m, (f) 500\,$\mu$m, (g) 710\,$\mu$m, and (h) 1000\,$\mu$m. }\label{fig:solidscontours}
\end{figure}

One of the most important aspects of the slurry flow behavior is the distribution of solid particles in radial direction. The solids concentration distribution influences other flow parameters and conditions such as pressure drop and velocity distribution directly and is of great importance in pipeline design \citep{embarka}. As discussed earlier, the solid particles in this study are polydisprese and have been considered as separate secondary phases in the mixture model. This approach makes the investigation of concentration distribution for solid particle with specific sizes possible. In Fig.\,\ref{fig:solidscontours}, the contours of concentration distribution of solid particles are shown for Case 2 from the smallest to largest particles, respectively. As depicted in Fig.\,\ref{fig:solidscontours}a\textendash h, the particles with smaller sizes (75 to 250\,$\mu$m) are more dispersed in the flow while larger particles (Fig.\,\ref{fig:solidscontours}e\textendash h) tend to settle and accumulate at the pipe invert. This observation is consistent with the experimental measurements of solids concentration distribution for multisized zinc particles through a lab\textendash scale pipe reported by \citet{KAUSHAL2002}. This trend in solids distribution is mainly due to gravitational effect, which makes the larger particles with higher weights to settle, but smaller particles are more prone to dispersion owing to lower weights. Based on these results, the pipe in the cross\textendash sectional view can be divided into three zones as explained by \citet{Liu2019}. The bottom section is the first zone where the larger particles settle, the second zone is the central region where the flow is steady and the particles are transported with the flow and are smaller in size, and the top part is the third zone where the small particles are suspended.

\section{Parametric study}

In this section, a systematic parametric study is presented to investigate the effect of various flow conditions on the behavior of the three\textendash phase slurry flow. The effects of particles size distribution, mixture velocity, Casson viscosity, and pipe angle on the pressure drop, solids concentration distribution, carrier velocity distribution, turbulent kinetic energy, and bitumen concentration distribution were studied. The solids concentration distribution and pressure drop determine the specific energy consumption (SEC) for the pipeline, which is a critical economic and environmental parameter. A lower SEC is favorable, and leads to transportation of more solid particles with a lower energy consumption. The magnitude and distribution of turbulent kinetic energy was investigated to evaluate the strength and regime of turbulence in the flow. 

For this parametric study, a base case with the flow conditions described as Case 2 in Table\,\ref{tab:expconditions} was chosen. The particles size distribution information was provided in Fig.\,\ref{fig:psd-density}a (Set1). To understand the role of each parameter on the flow behavior, other flow conditions were kept constant. The results demonstrate the nontrivial effect of aforementioned parameters on the flow behavior of the described slurries.  
\subsection{Effect of mixture velocity}

The mixture velocity has been proven to have a significant effect on the flow behavior of slurries \citep{sadeghi}. In this section, the effect of mixture velocity has been investigated by varying the flow velocity from 3.5\,m/s to 6.5\,m/s. In the real process which has been used for validation, the mixture velocity is in a range of 5.1 to 5.7\,m/s, which leads to a highly turbulent flow considering the pipe diameter. Reducing the flow velocity to 3.5\,m/s can give an idea about the flow behavior in a weak turbulent regime, and by increasing the velocity, the role of a stronger turbulence can be better understood in a three\textendash phase flow with a Casson carrier fluid. 

\begin{figure}
\centering
\centering
  \includegraphics[width=0.99\textwidth]{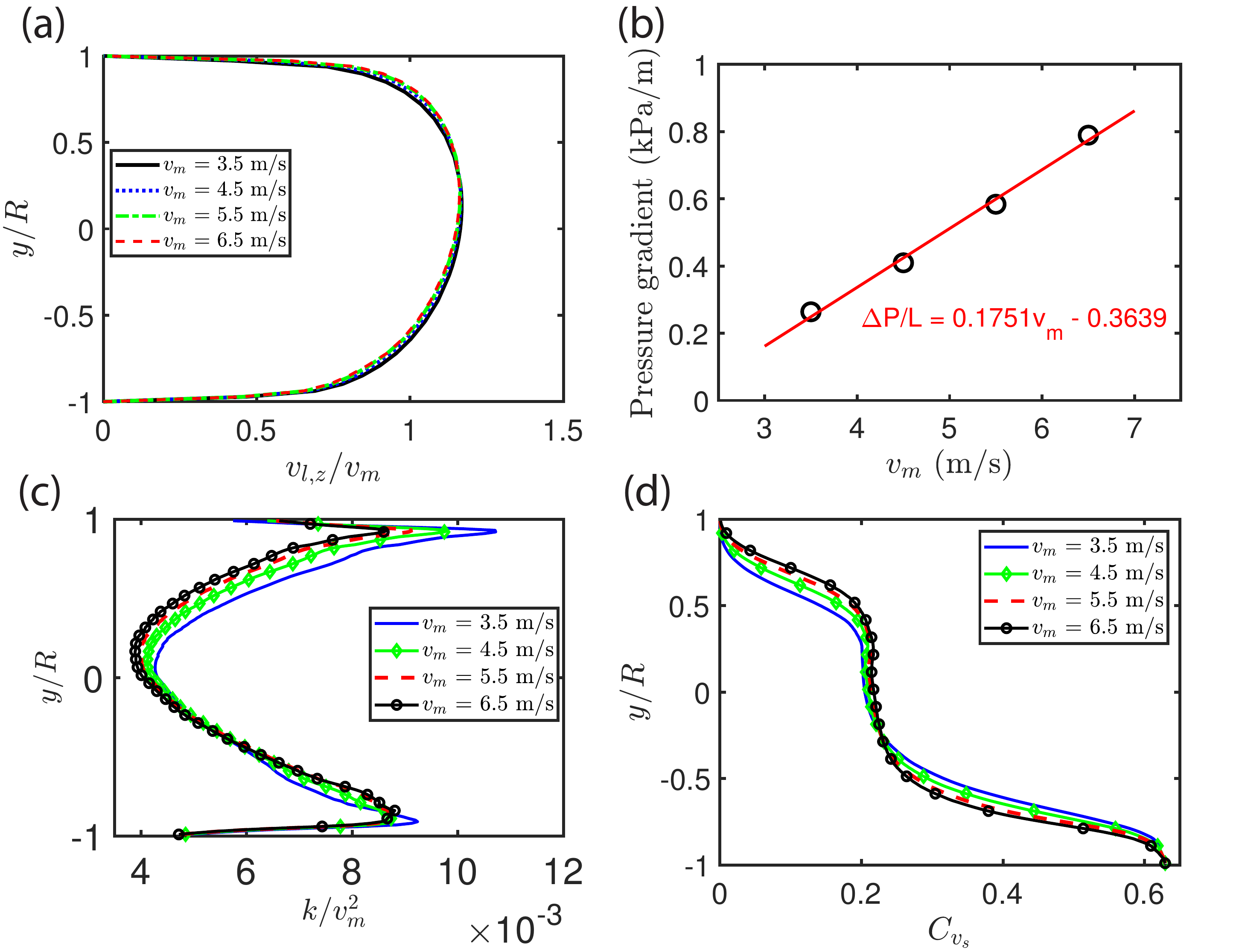}
\caption{Effect of mixture velocity on (a) carrier fluid velocity distribution, (b) pressure gradient, (c) chord\textendash averaged turbulent kinetic energy, and (d) chord\textendash averaged total solids concentration. (\begin{math}D = 74.0\,\mathrm{cm}, \alpha_v = 0.23, \rho_s = 2650\,\mathrm{kg/m^3},  \rho_l = 1278\,\mathrm{kg/m^3},
\tau_y = 0.04696\, \mathrm{Pa}, \mu_c = 0.00165\,\mathrm{Pa^{1/2}.s^{1/2}}) \end{math} }\label{fig:parametric-vm}
\end{figure}

In Fig.\,\ref{fig:parametric-vm}\,a, the normalized carrier velocity distributions in the vertical centerline have been plotted for the four simulated cases. While the curves of different mixture velocities seem to overlap, a slight shift of the velocity distribution toward a more asymmetrical form is observable with increasing the mixture velocity. The similar normalized velocity distribution demonstrates that the flows for all of the cases are in a similar regime. Similar trend is seen in the work by \citet{zhang2021} but for the solid phase velocity distribution, where the mixture velocity seems to have a trivial effect on the normalized velocity compared to other parameters. 

The pressure drop shows a significant dependency on mixture velocity as depicted in Fig.\,\ref{fig:parametric-vm}\,b. As the mixture velocity increases and the flow becomes more turbulent, the particle\textendash particle and particle\textendash wall collisions are enhanced significantly, leading to an intensified pressure drop in the pipeline. The pressure drop shows an increase of 200\,\% from an increase in the mixture velocity of 3.5 to 6.5\,m/s. The increase of pressure drop with mixture velocity is fairly linear as shown in this figure. Scaling equation is proposed to predict the pressure drop with any mixture velocity in the turbulent regime. By considering the specific energy consumption for the simulated cases, the pressure drop is increasing with mixture velocity while the delivered concentration is almost the same. It indicates that more power is required for the pumps to operate the transportation per delivered solids. 

The plots of nondimensionalized turbulent kinetic energy distribution are presented in Fig.\,\ref{fig:parametric-vm}\,c. CFD predictions demonstrate that the distribution of TKE is almost in the same form for all cases. Although increasing the mixture velocity intensifies turbulence, the flow regime seems to be similar which is accord with the velocity distributions shown in Fig.\,\ref{fig:parametric-vm}\,a. The lower TKE at the top region of the pipe for the cases with higher mixture velocities is probably due to presence of more solid particles dispersed in the flow.

Fig.\,\ref{fig:parametric-vm}d shows the total solids concentration distribution in radial direction. From the lowest mixture velocity to the highest one, a gradual decrease in the solids concentration in the lower region of the pipe is seen. This shows that more solid particles are suspended owing to the intensified turbulent regime and its mixing effects, but still turbulence is not able to fully suspend the large particles settled on the pipe invert. Particle\textendash particle interactions also contribute to higher suspension of solids at higher velocities, as these interactions are enhanced with increasing the velocity \citep{embarka}. These findings are in accord with the results achieved by \citet{akaushal2012} for a two\textendash phase slurry flow with fine particles, they mention that at the higher mixture velocity solid particles tend to migrate to the central part of the pipe due to slip velocity, and higher mixing effect from turbulence. Similar trend was also reported by \citet{Li2018ocean} from both experimental and simulation results with the mixture model, where the concentration of solid particles decreases near the pipe invert, and the asymmetry of the concentration curve is reduced at higher velocities.

The effect of mixture velocity on the cross\textendash sectional bitumen concentration distribution is presented in Table\,\ref{tab:bitdistribution}. As the mixture velocity increases, the droplets are more dispersed in the flow leading to a decline in the fraction of bitumen droplets at the top 25 one\textendash fourth and half of the pipe. More specifically, the ratio of bitumen droplets present at the top 25\,\% of the pipe to the total droplets decreases from 48.96\,\% for the case with $v_m$ = 3.5\,m/s to 41.13\,\% for the case with the highest mixture velocity of $v_m$ = 6.5\,m/s. This change is attributed to the higher mixing as a results of stronger turbulence as the velocity rises, similar to the effect of mixture velocity on the total solids concentration distribution discussed above.

\newpage
\subsection{Effect of particles size distribution (PSD)}

\begin{figure}
\centering

\includegraphics[width=0.9\textwidth]{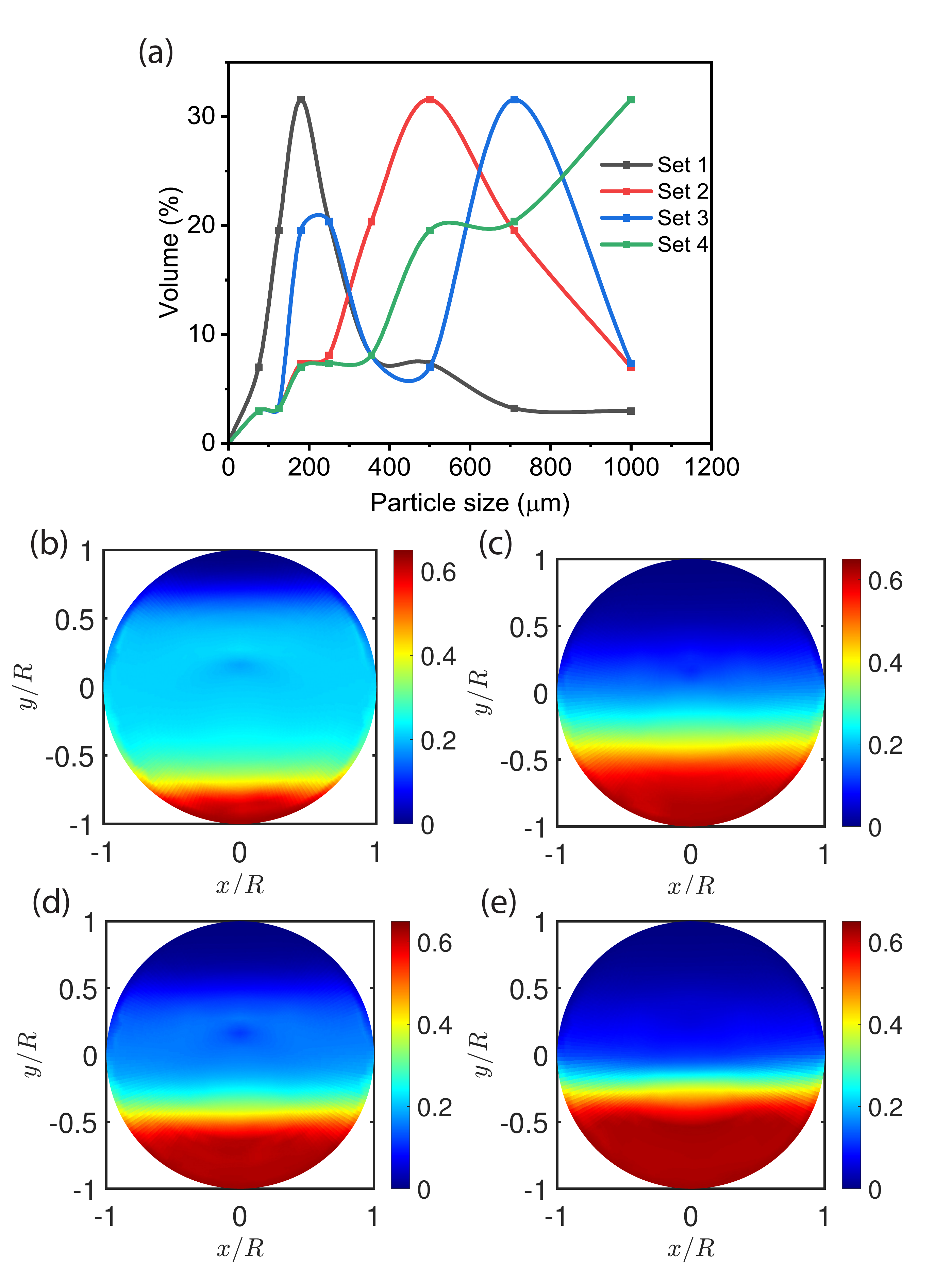}
  
\caption{[Color] (a) Plot of PSD information for the simulated cases to investigate the effect of PSD, Contours of total solids concentration distribution for (b) Set 1, (c) Set 2, (d) Set 3, and (e) Set 4 cases. (\begin{math}D = 74.0\,\mathrm{cm}, \alpha_v = 0.23, v_m = 5.35\,m/s, \rho_s = 2650\,\mathrm{kg/m^3},  \rho_l = 1278\,\mathrm{kg/m^3},
\tau_y = 0.04696\, \mathrm{Pa}, \mu_c = 0.00165\,\mathrm{Pa^{1/2}.s^{1/2}}) \end{math}}\label{fig:psd}
\end{figure}

In this section, the effect of PSD is investigated, by keeping the total solids concentration constant and varying the fraction of solid phases (corresponding to different sizes). Fig.\,\ref{fig:psd}a depicts PSD information for the four cases simulated for this section. Set\,1 represents the base case of the parametric studies, other sets are a result of shifting the maximum fraction to larger solid sizes. Set\,4 represents the case with the highest fraction of large solid particles, i.e., $d_p>$ 500\,$\mu m$.

Fig.\,\ref{fig:psd}b\textendash e show the contours of solids concentration distribution for Set 1 to Set 4 cases, respectively. As the concentration of larger particles increases, more particles tend to settle at the pipe invert, and form a sliding bed. Fig.\,\ref{fig:psd} confirms this argument, showing a higher solids concentration at the pipe bottom and a wide gap between the solids concentration at the top and bottom regions for the Set\,4 case (panel e).

\begin{figure}
\centering
\centering
 \includegraphics[width=0.9\textwidth]{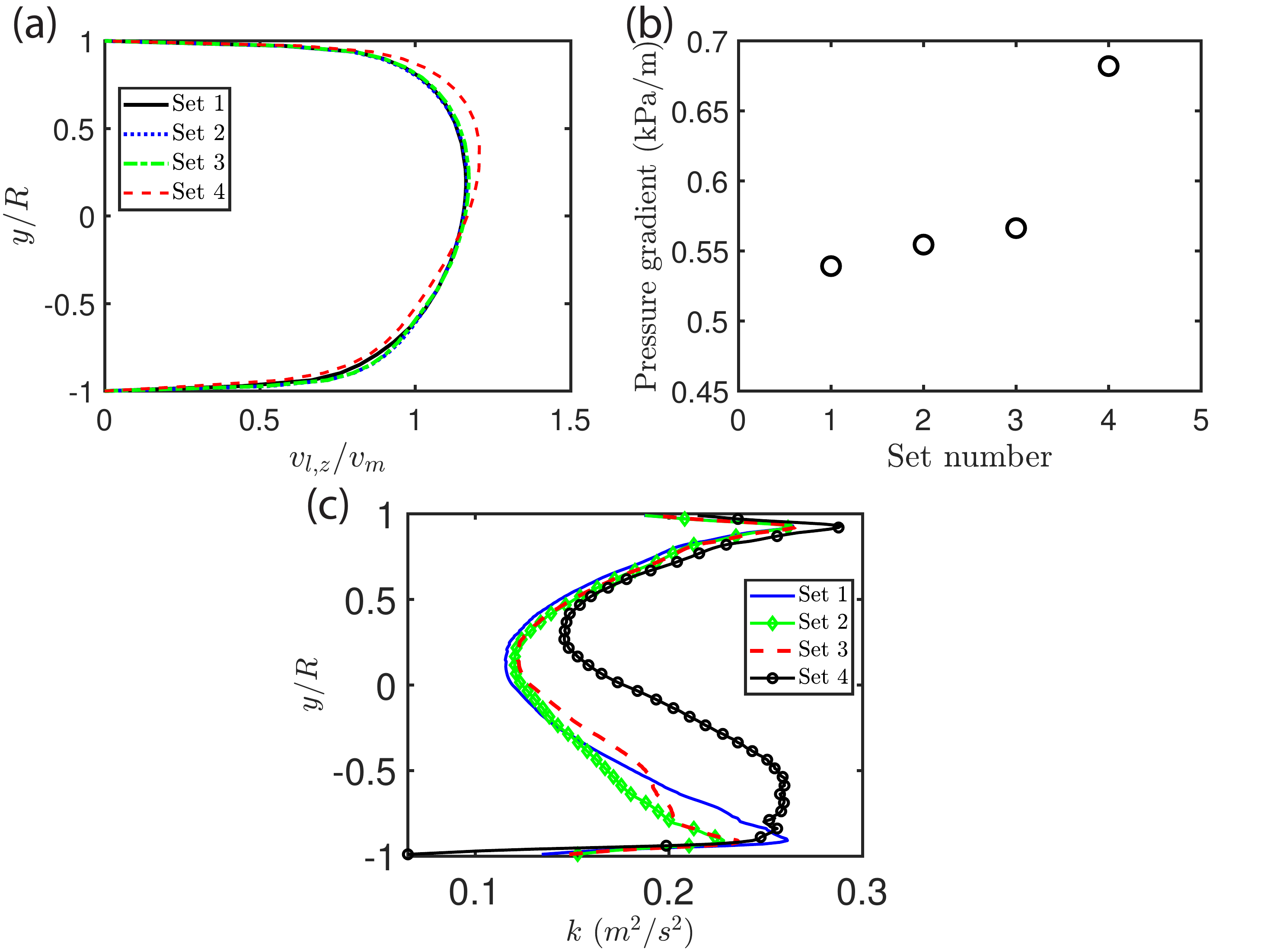}
\caption{Effect of PSD on (a) carrier fluid velocity distribution, (b) pressure gradient, and (c) chord\textendash averaged turbulent kinetic energy. (\begin{math}D = 74.0\,\mathrm{cm}, \alpha_v = 0.23, v_m = 5.35\,m/s, \rho_s = 2650\,\mathrm{kg/m^3},  \rho_l = 1278\,\mathrm{kg/m^3},
\tau_y = 0.04696\, \mathrm{Pa}, \mu_c = 0.00165\,\mathrm{Pa^{1/2}.s^{1/2}}) \end{math} }\label{fig:parametric-psd}
\end{figure}

Fig.\,\ref{fig:parametric-psd}\,a shows the carrier velocity distribution at the vertical centerline. The carrier velocity distribution is directly dependent on the solids concentration distribution \citep{embarka}. The carrier velocity declines near the bottom and shows an increase near the top of the pipe when the fraction of larger sizes increases\textendash see the red curve in Fig.\,\ref{fig:parametric-psd}\,a.  The formation of the bed leads to a decline in the effective flow area and in the mixture and consequently carrier velocity at the bottom region. Contrariwise, less solid particles are present at the top region, and the carrier can freely flow in this region, with having a higher velocity. \citet{zhang2021} reported the similar results for the monodisperse solid particles, where by increasing the size the maximum velocity occurs at a higher vertical position.

The effect of PSD on the frictional pressure drop in the pipe is presented in Fig.\,\ref{fig:parametric-psd}\,b. The pressure drop increases with shifting the maximum fraction to larger particle sizes, with showing a significant rise from Set\,3 to Set\,4. This increase in pressure drop was expected and is in accord with other results shown in this section. More particles settlements and particle\textendash wall collisions, also intensified particle\textendash particle collisions due to the lower distance between the particles lead to more energy loss and pressure drop. As mentioned before, the pressure drop in a slurry flow is mostly a result of particle\textendash particle and particle\textendash wall interactions, both intensified for the cases with higher fractions of large solid particles. From Set 1 to Set 3 case, the pressure gradient shows a 5\% increase, while from Set 3 to Set 4, a 20\% shift is seen in the value of pressure gradient. The effect of PSD on the pressure gradient is similar to the effect of particle size in a two\textendash phase flow with monodisperse particles. Several researchers have studied the effect of particle size on the flow behavior, and reported an increase in the pressure gradient as the particle size is raised \citep{gopaliya2015,zhang2021,sadeghi}. It must be noted that the trend and magnitude of changes in the pressure gradient with respect to PSD is different from slurries with single\textendash sized particles.

In Fig.\,\ref{fig:parametric-psd}\,c, the chord\textendash averaged turbulent kinetic energy is presented to evaluate the effect of PSD on turbulent regime and velocity fluctuations. The results show that for all of the cases, the turbulent kinetic energy shows a minimum at the pipe center, and local maximums in the bottom half of the pipe. For the fourth case (Set\,4), the turbulent kinetic energy is approaching to zero near the wall, which is due to a dense solid bed at the pipe invert, damping the turbulence dramatically. This is consistent with the analysis presented by \citet{antaya} and \citet{Gopaliya2016}, which show a lower magnitude in the turbulence modulation where the solids concentration is higher.  However, for the top region of the pipe in the Set 4 case, the turbulent kinetic energy is higher the other cases. This shows that for this case, turbulent is stronger and more active in this region, constantly changing the direction of the velocity vectors. This is in accord with the velocity distribution depicted in Fig.\,\ref{fig:parametric-psd}\,a which shows higher velocities at the top region for Set 4.   

The effect of PSD on the bitumen concentration distribution is quantitatively analyzed in Table \ref{tab:bitdistribution}. As shown, with increasing the fraction of larger sizes from Set 1 to Set 4, a slight increase in the bitumen concentration at the top half of the pipe is seen, which is due to less small particles present in the flow to occupy that region along bitumen. However, the difference for the top 25\,\% is less pronounced, from 43.08\,\% for Set 1 to 43.7\,\% for Set 4 case. Overall, the bitumen concentration distribution in the vertical direction seem to be insignificantly dependent on PSD along with the other parameters.

\newpage
\subsection{Effect of Casson viscosity}

\begin{figure}
\centering
 \includegraphics[width=0.9\textwidth,height=0.9\textheight,keepaspectratio]{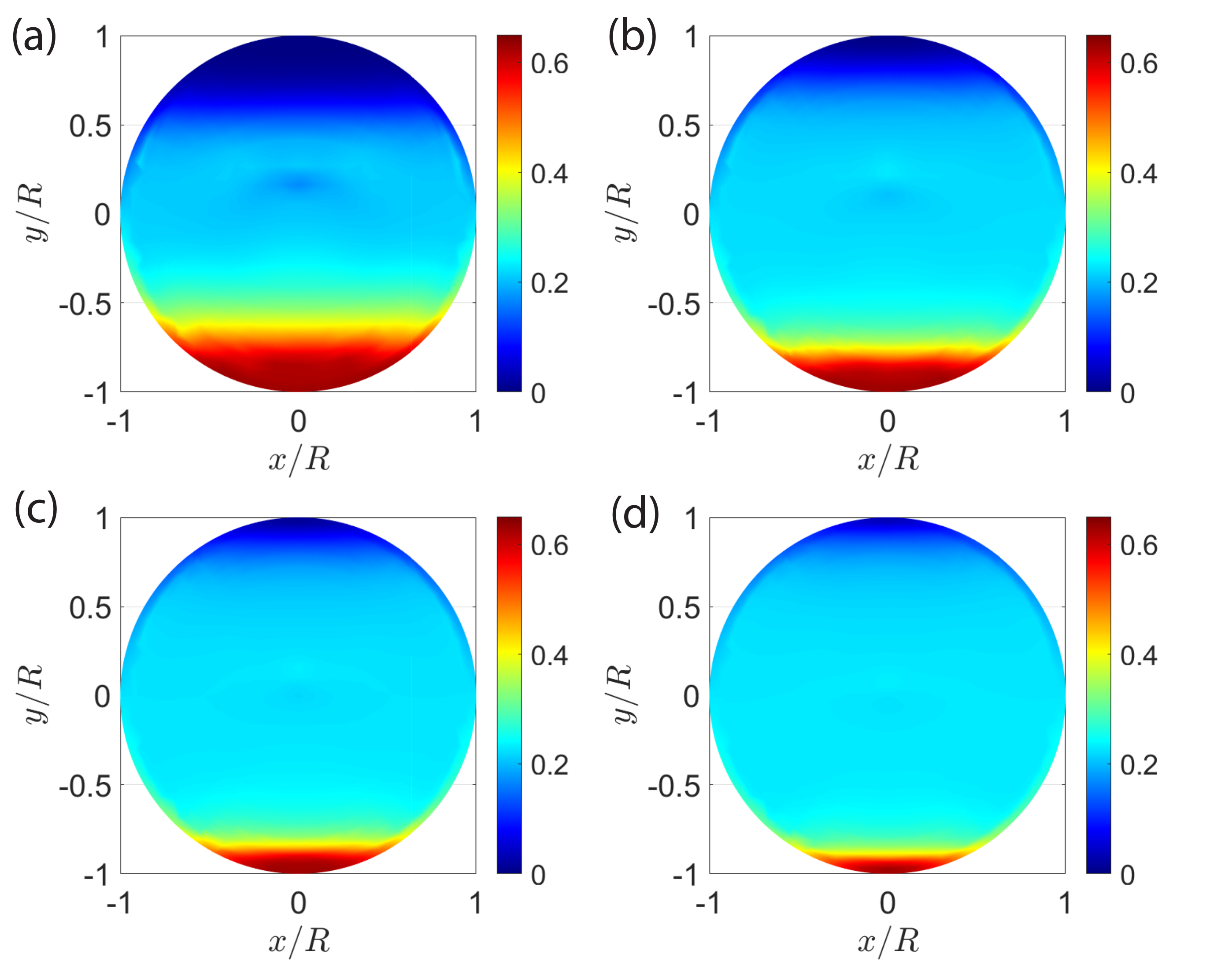}
\caption{[Color] Contours of total solids concentration distribution for the cases with Casson viscosity of (a) 0.001 $Pa^{1/2}.s^{1/2}$, (b) 0.003 $Pa^{1/2}.s^{1/2}$, (c) 0.006 $Pa^{1/2}.s^{1/2}$, and (d) 0.01 $Pa^{1/2}.s^{1/2}$. (\begin{math}D = 74.0\,\mathrm{cm}, \alpha_v = 0.23, v_m = 5.35\,m/s, \rho_s = 2650\,\mathrm{kg/m^3},  \rho_l = 1278\,\mathrm{kg/m^3},
\tau_y = 0.04696\, \mathrm{Pa}) \end{math}}\label{fig:casson_solidvof}
\end{figure}

Fig.\,\ref{fig:casson_solidvof} shows the effect of Casson viscosity on the solids concentration distribution in radial direction inside the pipe. The results show that a higher Casson viscosity results in more suspension of the particles in the carrier fluid (Fig.\,\ref{fig:casson_solidvof}\,d). With increasing Casson viscosity, there is a consistent decrease in total solids concentration at the pipe invert. This observation could be due to the more viscous carrier fluid, having a higher core viscosity which leads to more support of the particles in the flow and more suspension. Casson plastic viscosity quantifies the internal frictions of a fluid, and it is proven that a more viscous carrier fluid can support more particles in the core region and prohibit the stratification of the flow. 

\begin{figure}
\centering
\centering
  \includegraphics[width=0.9\textwidth,height=0.9\textheight,keepaspectratio]{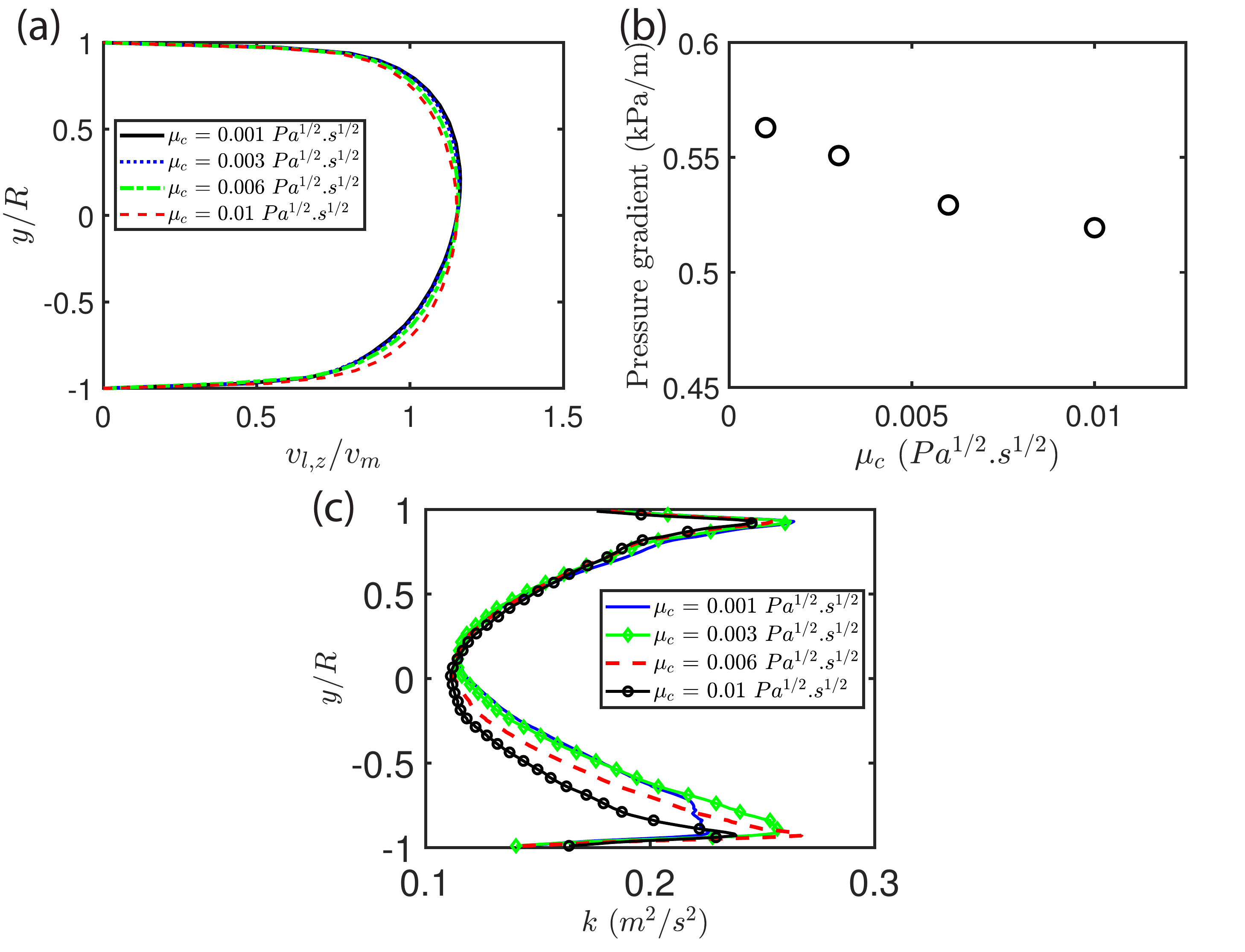}
\caption{Effect of Casson Viscosity on (a) carrier fluid velocity distribution, (b) pressure gradient, and (c) chord\textendash averaged turbulent kinetic energy. (\begin{math}D = 74.0\,\mathrm{cm}, \alpha_v = 0.23, v_m = 5.35\,m/s, \rho_s = 2650\,\mathrm{kg/m^3},  \rho_l = 127\,\mathrm{kg/m^3},
\tau_y = 0.04696\, \mathrm{Pa}) \end{math}}\label{fig:parametric-u}
\end{figure}

The carrier velocity distributions for the simulated cases are presented in Fig.\,\ref{fig:parametric-u}\,a. For case with a higher Casson viscosity, the velocity distribution is more symmetric. In case of a higher viscosity and constant mixture velocity, the Reynolds number decreases as the Casson viscosity rises, resulting in a weaker turbulence in the system. The weaker turbulence in a flow with the higher Casson viscosity is also identified from plots of turbulent kinetic energy.

The pressure gradient for the flows with different Casson viscosity for the carrier fluid was compared as shown in Fig.\,\ref{fig:parametric-u}\,b. The results show that the increase in Casson viscosity leads to a lower pressure gradient, however, the changes in pressure gradient are not dramatic. From the case with $\mu_c$ = 0.001 to $\mu_c$ = 0.01\,$Pa^{1/2}.s^{1/2}$, a decrease of 7\% is identified on pressure gradient. Additionally, the pressure gradient changes are not uniform, with less significant changes occurring as Casson viscosity increases. The interactions between the wall and particles are dampen as the solid particles are more suspended in the flow. On the other hand, the internal frictions are enhanced as the Casson viscosity is raised, which act against each other in affecting the pressure drop. Furthermore, the effect of weakened particle\textendash wall collisions are more effective in reducing the pressure compared to the increase in pressure drop due to more viscous fluid. 

In Fig.\,\ref{fig:parametric-u}\,c, the radial distributions of turbulent kinetic energy for the four cases are depicted, showing an inverse proportion especially in the core region of the pipe between the turbulent kinetic energy and Casson viscosity. The relatively lower turbulent kinetic energy for $\mu_c$ = 0.01 case at the pipe bottom half may be due to the presence of more solid particles in that region for this case (see Fig.\,\ref{fig:casson_solidvof}\,d) similar to our previous studied parameters. Also as mentioned, turbulence is weaker for the cases with higher Casson viscosity. 

These results are consistent with the analysis performed by \citet{enzu2021} on the effect of flow index (n) in a Herschel\textendash Bulkley carrier fluid on the flow behavior of a two\textendash phase slurry. In their study, by increasing the value of flow index, the flow shifts toward a laminar flow, leading to a more uniform solid distribution, lower values for turbulent kinetic energy, and a change in the velocity profile from a turbulent to a laminar flow with a more symmetrical curve.

In the fourth section of Table \ref{tab:bitdistribution}, the ratio of the bitumen droplets present at the top 25 and 50\,\% of the pipe is shown. The results demonstrate that the bitumen ratio at the top region is inversely proportional with Casson viscosity. When the value of Casson viscosity is raised from 0.001 to 0.01\,$Pa^{1/2}.s^{1/2}$, the bitumen ration decreases from 44.26 to 40.31\,\% for the top one\textendash fourth, and from 70.8 to 63.87\,\% for the top half of the pipe. There is a decrease in the bitumen ratio due to the coexistence of more solid particles with the bitumen droplets at the top region for the case with the highest Casson viscosity, as shown in Fig\,\ref{fig:casson_solidvof}.

\subsection{Effect of pipe angle}

 In this section, two cases for the inclination with angles of +5 and +10° and two cases for declination with angles of -5 and -10° are simulated. Other pipe characteristics such as diameter and length also the flow conditions are kept constant in these cases.

\begin{figure}
\centering
 \includegraphics[width=0.9\textwidth,height=0.9\textheight,keepaspectratio]{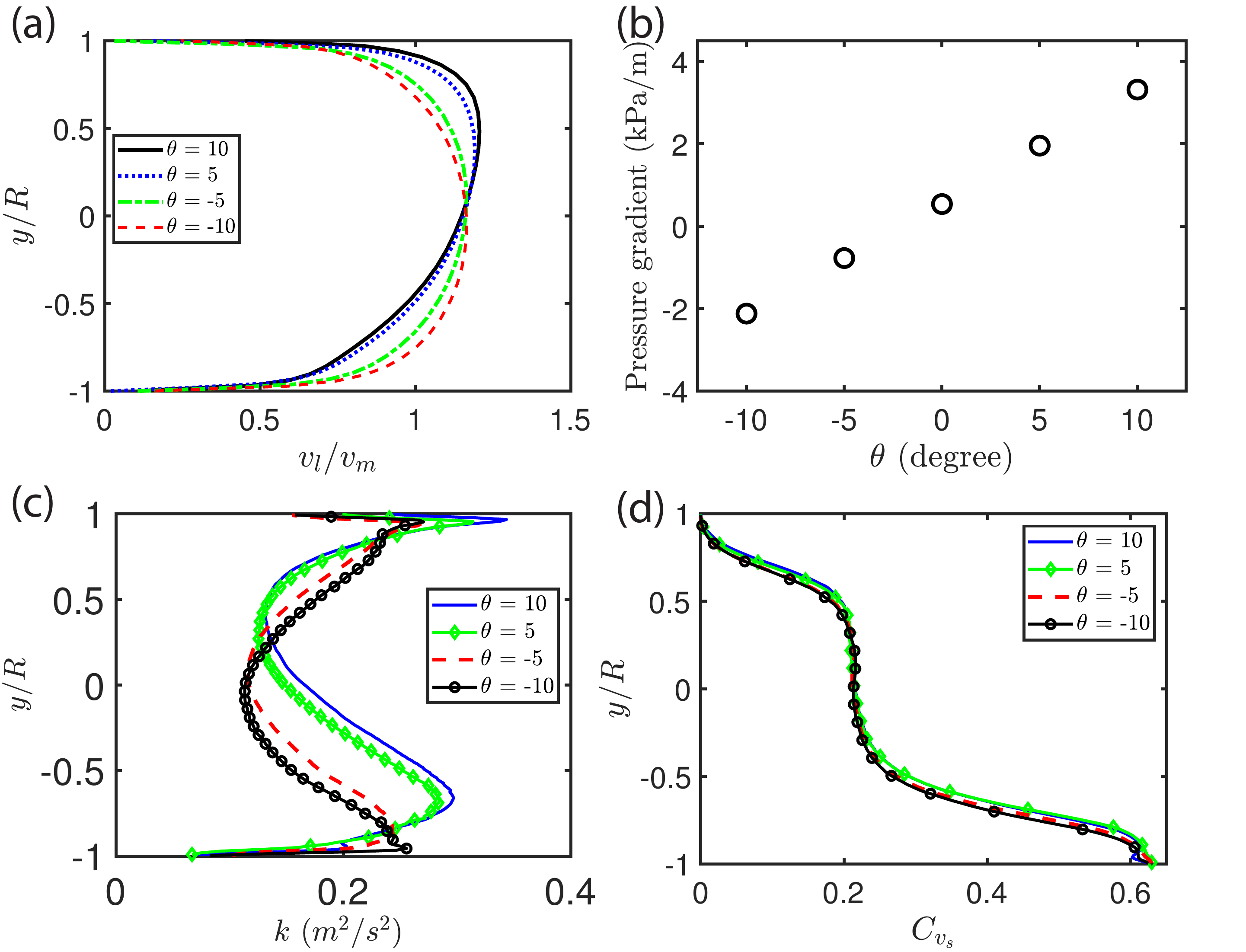}
\caption{Effect of pipe angle on (a) carrier fluid velocity distribution in the $z$ direction, (b) pressure gradient, (c) chord\textendash averaged turbulent kinetic energy, (d) chord\textendash averaged total solids concentration. (\begin{math}D = 74.0\,\mathrm{cm}, \alpha_v = 0.23, v_m = 5.35\,m/s, \rho_s = 2650\,\mathrm{kg/m^3},  \rho_l = 1278\,\mathrm{kg/m^3},
\tau_y = 0.04696\, \mathrm{Pa}, \mu_c = 0.00165\,\mathrm{Pa^{1/2}.s^{1/2}}) \end{math} }\label{fig:parametric-ang}
\end{figure}

Fig.\,\ref{fig:parametric-ang}a shows the carrier velocity distribution in z\textendash direction. The pipe angle has a significant effect on the velocity distribution as the effect of gravity is changing in the cases with the angle (in y\textendash direction). In an inclined pipe, the flow has a higher velocity at the upper half of the pipe, while the velocity distribution is fairly symmetrical in case of a declined pipe. The effect of pipe angle on the pressure drop is shown in Fig.\,\ref{fig:parametric-ang}b. The results demonstrate that the pressure drop has a linear relationship with the pipe angle. It must be noted that the negative sign of the pressure drop in the declined cases comes from the zero gauge pressure defined at the pipe outlet as a boundary condition. The experimental and model results achieved by \citet{DORON1997313} confirm this trend in the pressure drop with the pipe tilt. In case of the declined pipe, the gravity acts as an driving force for the flow and can eliminate the energy required from the pumps if the slope is high enough.

The pipe angle was proven to have also a nontrivial effect on the turbulence regime as shown with the analysis of chord\textendash averaged turbulent kinetic energy in the radial direction in Fig.\,\ref{fig:parametric-ang}c. As shown in Fig.\,\ref{fig:parametric-ang}c, there are more significant velocity fluctuations and consequently stronger turbulence at the top region for the declined cases as well as at the bottom for the inclined cases. As shown in Fig.\,\ref{fig:parametric-ang}d, solids concentration distribution is weakly dependent on the pipe angle, with slight differences at the pipe obvert and invert. This observation is in accord with the experimental results reported by \citet{kesely2019} and \citet{Vlasak2020} for two\textendash phase slurries in declined and inclined pipes, which shows when the difference in the pipe angle is not large (less than 20°), the particles distribution is similar. The tendency of solid particles to settle is slightly higher in inclined pipes, which is in accord with previously shown results for velocity distribution and pressure drop. Since the effective flow area of inclined pipes is smaller due to more solid particles at the invert, velocities at the top are higher, while enhanced particle\textendash wall particle\textendash wall interactions result in higher pressure drops.

As shown in Table\,\ref{tab:bitdistribution}, the pipe angle has a non\textendash uniform effect on the bitumen concentration distribution. In the case of shifting from the negative values of the pipe angle to the positive values of the pipe angle (inclined pipes), the bitumen ratio increases at first, then declines further as the angle increases. However, the magnitude of the changes in bitumen concentration distribution with the pipe angle is insignificant and can almost be ignored.

\newpage
\section{Conclusions}

We developed an efficient three-dimensional computational CFD model to study the flow behavior of a turbulent three-phase non-Newtonian tailings slurry flow in an industrial pipeline. The Gidaspow drag model was selected for carrier\textendash solids, and the sub\textendash models for KTGF were selected from the tested standard models. The SST k\textendash $\omega$ turbulence model was used to model highly turbulent flow in a pipeline. The Casson rheological model coupled to capture the non\textendash Newtonian behavior of the carrier fluid in a slurry system. The accuracy of the model was proved by validating the results against collected filed data. The CFD results showed exceptional agreement with the field data, with errors of $<$\,3.5\% for velocity distribution and $<$\,15\% for the pressure drop. We further extended our CFD model applicability to study a comprehensive parametric study to investigate the influence of mixture velocity, particle size distribution, carrier fluid rheology, and pipe angle on the behavior of the slurry flow, including bitumen concentration, solids concentration, and pressure drop. The results of the CFD demonstrated that most bitumen droplets reside in the top region of the pipe and selective treatment of the top region may lead to an acceptable bitumen recovery. The CFD predictions on the bitumen concentration distribution would be helpful for the separation of bitumen residues. The developed CFD model provides a powerful tool to predict flow behaviors during highly turbulent slurry transport and may guide the process design for bitumen recovery on an industrial scale.

\section*{Credit author statement}
\textbf{Mohsen Sadeghi:} Conceptualization, Data curation, Methodology, Investigation, Validation, Formal analysis, Writing\textendash original draft, Writing\textendash review \& editing. \textbf{Somasekhara Goud Sontti:} Conceptualization, Methodology, Investigation, Validation, Writing\textendash review \& editing. \textbf{Enzu Zheng:} Writing\textendash review \& editing, Validation. \textbf{Xuehua Zhang:} Conceptualization, Methodology, Supervision, Project administration, Resources,Writing\textendash review \& editing.
\section*{Declaration of competing interest}
The authors declare that they have no known competing financial interests or personal relationships that could have appeared to influence the work reported in this paper.

\section*{Acknowledgments}
The authors acknowledge the funding support from
the Institute for Oil Sands Innovation (IOSI) (Project number IOSI 2019\textendash 04 (TA)) and from the Natural Science and Engineering Research Council of Canada (NSERC)\textendash Alliance. This research was undertaken, in part, thanks to funding from the Canada Research Chairs Program. We also thank Compute Canada (www.computecanada.ca) for continued support through extensive access to the Compute Canada HPC Cedar and Graham clusters. The authors are grateful to Dr.\,Petr Nikrityuk for the valuable advice. 


\mbox{}

\nomenclature[A]{$D$}{pipe diameter (L)}
\nomenclature[A]{$R$}{pipe radius (L)}
\nomenclature[A]{$d_p$}{particle diameter (L)}
\nomenclature[A]{$C_{v}$}{chord\textendash averaged concentration (--)}
\nomenclature[A]{$g$}{gravitational acceleration (L T$^{-2}$)}
\nomenclature[A]{$g$}{acceleration (L T$^{-2}$)}
\nomenclature[A]{$g_0$}{radial distribution function (--)}
\nomenclature[A]{$p$}{locally\textendash averaged pressure (M L$^{-1}$ T$^{-2}$)}
\nomenclature[A]{$t$}{time (T)}
\nomenclature[A]{$v$}{velocity (L T$^{-1}$)}
\nomenclature[A]{$V$}{velocity (L T$^{-1}$)}
\nomenclature[A]{$f_\mathrm{drag}$}{drag function (--)}

\nomenclature[A]{$C_\mathrm{fr}$}{friction coefficient between solid phases (--)}
\nomenclature[A]{$x$}{horizontal coordinate (L)}
\nomenclature[A]{$y$}{vertical coordinate (L)}
\nomenclature[A]{$z$}{axial coordinate (L)}
\nomenclature[A]{$e$}{restitution coefficient (--)}
\nomenclature[A]{$I_{2D}$}{second invariant of the deviatoric stress tensor (--)}
\nomenclature[A]{$\norm{\vec{v}_s^{\,\prime}}$}{fluctuating solids velocity (L T$^{-1}$)}
\nomenclature[A]{$i$}{hydraulic gradient (--)}
\nomenclature[A]{$K_{ls}$}{momentum exchange coefficient between fluid
\nomenclature[A]{$\Delta P$}{area\textendash averaged gauge pressure (M L$^{-1}$ T$^{-2}$))}
\nomenclature[A]{$k$}{turbulent kinetic energy (L$^{2}$ T$^{-2}$))}

\nomenclature[G]{$\alpha$}{locally\textendash averaged volume fraction (--)}
\nomenclature[G]{$\mu$}{dynamic viscosity (M L$^{-1}$ T$^{-1}$)}
\nomenclature[G]{$\rho$}{density (M L$^{-3}$)}
\nomenclature[G]{$\sigma_{t}$}{Prandtle/Schmidt number}
\nomenclature[G]{$\phi_{ls}$}{the energy exchange between the fluid and the solid phases (E) }
\nomenclature[G]{$\gamma_{\Theta_{s}}$}{collisional dissipation of energy (E)}
\nomenclature[G]{$\tau$}{shear stress (M L$^{-1}$ T$^{-2}$)}
\nomenclature[G]{$\dot{\gamma}$}{shear strain rate (T$^{-1}$)}
\nomenclature[G]{$\alpha_{s,\mathrm{max}}$}{maximum packing limit (--)}
\nomenclature[G]{$\Theta$}{granular temperature (L$^{-2}$ T$^{-2}$)}
\nomenclature[G]{$\varphi$}{angle of internal friction (--)}
\nomenclature[G]{$\eta_{t}$}{turbulent diffusivity (--)}
\nomenclature[G]{$\eta$}{apparent viscosity (M L$^{-1}$ T$^{-1}$)}

\nomenclature[S]{$l$}{liquid}
\nomenclature[S]{$s$}{solid}
\nomenclature[S]{$m$}{mixture}
\nomenclature[S]{$ss$}{solid particles}
\nomenclature[S]{$p$}{$p^{th}$ solid phase}
\nomenclature[S]{$q$}{$q^{th}$ solid phase}
\nomenclature[S]{col}{collisional part of viscosity}
\nomenclature[S]{kin}{kinetic part of viscosity}
\nomenclature[S]{fr}{frictional part of viscosity}
\nomenclature[S]{$y$}{yield stress}
\nomenclature[S]{$dr$}{drift}

\printnomenclature

\newpage
\bibliography{mybibfile}
\newpage
\section*{Graphical abstract}
\begin{figure}[h!]
	\centering
	\includegraphics[width=\textwidth]{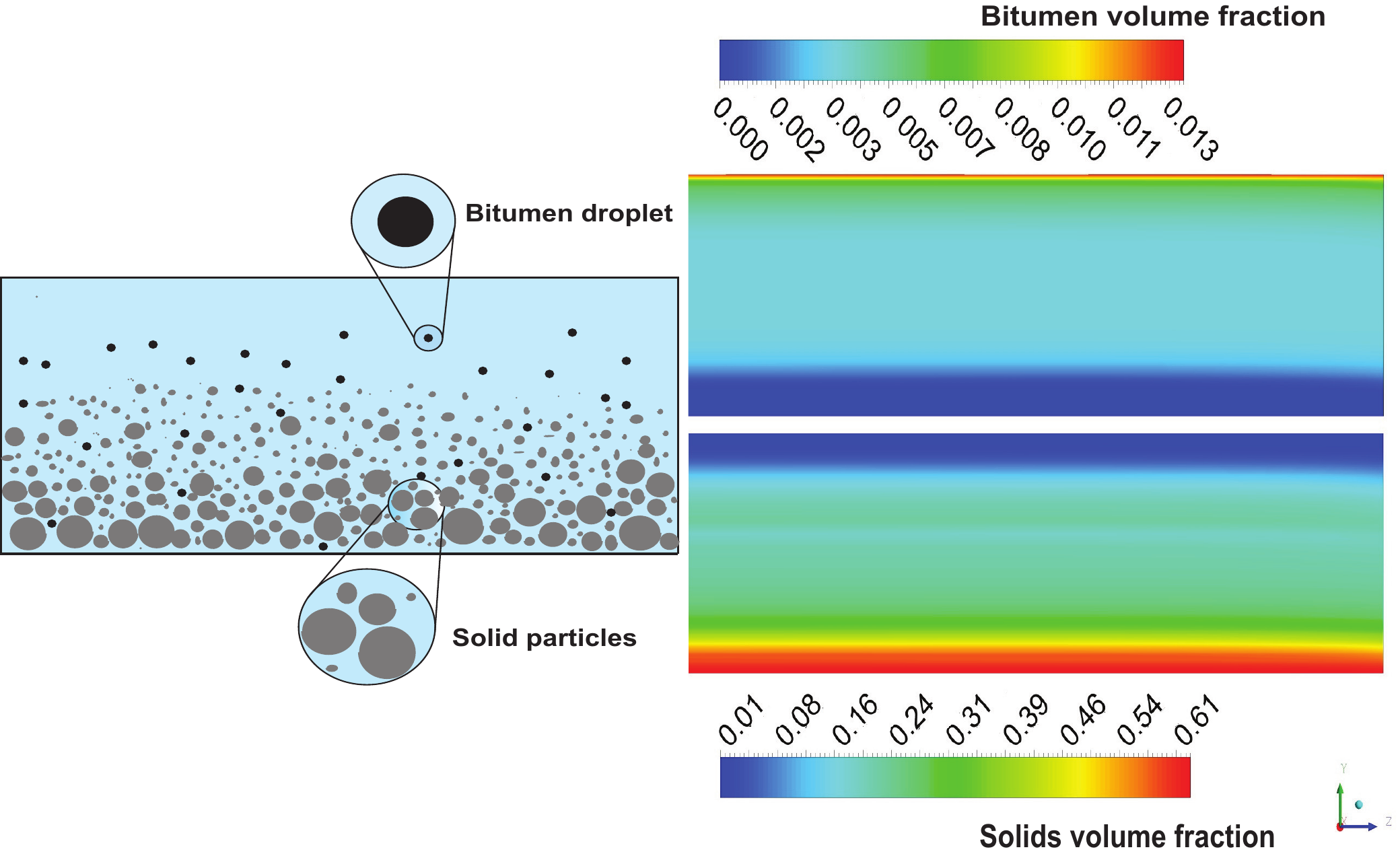}
\end{figure}
\section*{Highlights}

\begin{itemize}

\item	Developed an industrial-scale CFD model for turbulent tailings slurry in a pipeline

\item 	Good agreement in model validation with field data
\item Flow behaviors revealed in comprehensive parametric study
\item Predictions of preferential distribution of bitumen in the pipe 

\end{itemize}

\end{document}